\documentclass[sigconf]{acmart}
\usepackage[utf8]{inputenc}
\usepackage{graphicx}
\usepackage{subfigure}
\usepackage{booktabs}
\usepackage[disable]{todonotes} %
\usepackage{balance}

\usepackage[absolute,showboxes]{textpos}

\acmYear{2021}\copyrightyear{2021}
\setcopyright{acmcopyright}
\acmConference[IMC '21]{ACM Internet Measurement Conference}{November 2--4, 2021}{Virtual Event, USA}
\acmBooktitle{ACM Internet Measurement Conference (IMC '21), November 2--4, 2021, Virtual Event, USA}
\acmPrice{15.00}
\acmDOI{10.1145/3487552.3487835}
\acmISBN{978-1-4503-9129-0/21/11}


\newcommand{\done}[1]{}

\newcommand{\result}[1]{}

\definecolor{myred}{cmyk}{0, 0.7808, 0.4429, 0.1412}

\usepackage{pifont}

\usepackage{xspace}
\newcommand{\eg}{\textit{e.g.,}~}
\newcommand{\ie}{\textit{i.e.,}~}
\newcommand{\etc}{\textit{etc.}~}
\newcommand{\one}{({\em i})\xspace}
\newcommand{\two}{({\em ii})\xspace}
\newcommand{\three}{({\em iii})\xspace}

\makeatletter
\renewcommand{\paragraph}[1]{\vspace*{0.03in}\noindent{\bf #1.}\hspace{0.25ex \@plus1ex \@minus.2ex}}
\newcommand{\paragraphNoDot}[1]{\vspace*{0.03in}\noindent{\bf #1}\hspace{0.25ex \@plus1ex \@minus.2ex}}
\makeatother

\begin{document}

\date{}

\setlength{\TPHorizModule}{\paperwidth}
\setlength{\TPVertModule}{\paperheight}
\TPMargin{5pt}
\begin{textblock}{0.83}(0.08,0.02)
	\noindent
	\footnotesize
	\centering
	If you cite this paper, please use the IMC reference:
	M. Nawrocki, M. Jonker, T. C. Schmidt, and M. Wählisch.
	2021. The Far Side of DNS Amplification: Tracing the DDoS Attack Ecosystem from the Internet Core.
	\emph{In Proceedings of ACM Internet Measurement Conference (IMC ’21).}
	ACM, New York, NY, USA, 16 pages. https://doi.org/10.1145/3487552.3487835
\end{textblock}

\title[Tracing the DDoS Attack Ecosystem from the Internet Core]{The Far Side of DNS Amplification: Tracing the DDoS Attack Ecosystem from the Internet Core}

\author{Marcin Nawrocki}
\email{marcin.nawrocki@fu-berlin.de}
\affiliation{%
  \institution{Freie Universit{\"a}t Berlin}
  \country{Germany}
}

\author{Mattijs Jonker}
\email{m.jonker@utwente.nl}
\affiliation{%
  \institution{University of Twente}
  \country{Netherlands}
}

\author{Thomas C. Schmidt}
\email{t.schmidt@haw-hamburg.de}
\affiliation{%
  \institution{HAW Hamburg}
  \country{Germany}
}

\author{Matthias W\"ahlisch}
\email{m.waehlisch@fu-berlin.de}
\affiliation{%
  \institution{Freie Universit{\"a}t Berlin}
  \country{Germany}
}

\renewcommand{\shortauthors}{Nawrocki et al.}

\begin{abstract}

In this paper, we shed new light on the DNS amplification ecosystem, by
studying complementary data sources, bolstered by orthogonal methodologies.
First, we introduce a passive attack detection method for the Internet core,
\ie at Internet eXchange Points (IXPs).  Surprisingly, IXPs and
honeypots observe mostly disjoint sets of attacks: 96\% of IXP-inferred
attacks were invisible to a sizable honeypot platform.
Second, we assess the effectiveness of observed DNS attacks by studying IXP
traces jointly with diverse data from independent
measurement infrastructures.  We find that attackers efficiently detect
new reflectors and purposefully rotate between them. At the same time, 
we reveal that attackers are a small step away from bringing about significantly higher
amplification factors~($14\times$).
Third, we identify and fingerprint a major attack entity by studying patterns in attack traces. %
We show that this entity dominates the DNS amplification ecosystem by carrying out $59\%$ of the attacks, and provide an in-depth analysis of its behavior over time.
Finally, our results reveal that operators of various \texttt{.gov} names do not adhere to DNSSEC key rollover best practices, which exacerbates amplification potential.
We can verifiably connect this operational behavior to misuses and attacker decision-making.

\end{abstract}

\maketitle

\section{Introduction}

Denial of Service (DoS) attacks pose a major, omnipresent threat to the stability of the Internet.
About one-third of the active \texttt{/24} networks on the Internet
received DoS attacks over a two-year period~\cite{jonker2017millions}, and
90\% of attacks mitigated at a large IXP involved reflection
attacks~\cite{nawrocki2019down}.
To bring about reflection, attackers spoof source IP addresses to send
request packets that supposedly originate from an intended victim, and abuse
the infrastructure that replies to these requests (\eg open DNS resolvers).
Amplification is successful if the responses are larger than the requests.

The DNS is a core Internet component. It
primarily operates over the transport-layer protocol UDP. Due to its stateless
nature, UDP is particularly susceptible to spoofing, and at least 14 protocols
that work on top of UDP allow for reflection attacks~\cite{rossow2014amplification}.
The Network Time Protocol (NTP) and DNS are (currently) the most-abused protocols \cite{ddos2021kopp, nawrocki2019down, jonker2017millions}. %

Notably, amplification attacks are not limited to UDP.
Poor implementations of network stacks allow attackers to use TCP as well~\cite{kuhrer2014hell}.
A recent DNS amplification attack exploits inefficient resolver
implementations and works regardless of the underlying transport-layer
protocol~\cite{shafir2020nxnsattack}---DNS amplification remains one
of the most popular attack vectors, despite recent changes \mbox{such as
DNS-over-TLS~\cite{RFC-7858} and~DNS-over-HTTPS~\cite{RFC-8484}.}

Expert measurement methods are essential to observe global attack
activities.
Having a thorough understanding of attack dynamics
and the abused infrastructure is crucial to effectively mitigate DNS-based attacks and
to reduce the opportunity for infrastructure abuse.
Several efforts exist to monitor amplification attacks on
a global scale. Primarily, the monitoring infrastructures are
implemented with the help of honeypots~\cite{kramer2015amppot,
noroozian2016gets, thomas20171000}. In such works, careful assumptions are made about the share of global attacks
that honeypots account for~\cite{kramer2015amppot, thomas20171000} because the amplification ecosystem consists of a large number
of amplifiers~\cite{ShodanWebsite} with high churn
rates~\cite{kuhrer2014exit}. Moreover, sophisticated
attackers learn about the location of honeypots and exclude them~\cite{ph-vhfbt-08}.

\begin{figure}[t]
  \begin{center}
  \includegraphics[width=1\columnwidth]{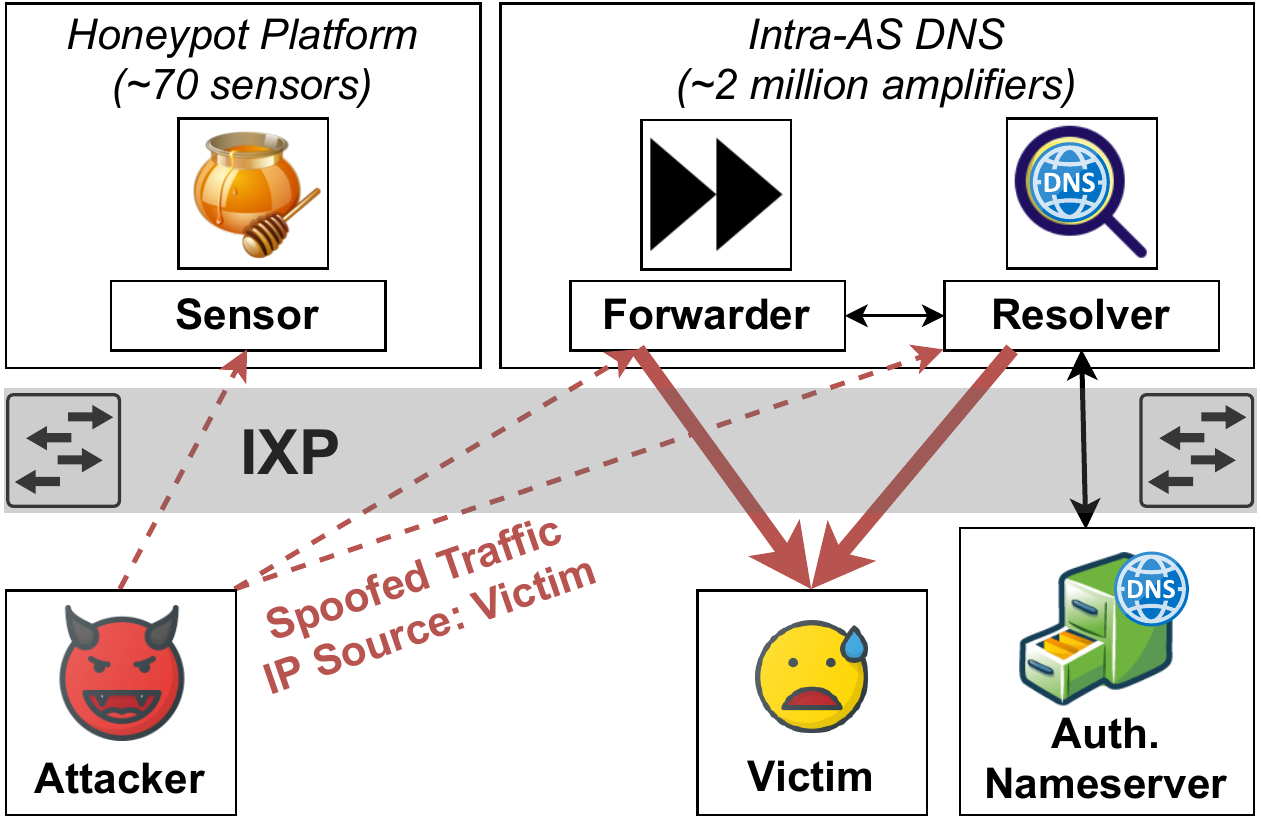}
      \caption{Vantage points and stakeholders of distributed, inter-domain DNS amplification attacks.}
  \label{fig:data_sketch}
  \end{center}
\end{figure}

In this paper, we extend the understanding of the DNS amplification ecosystem
by jointly analyzing results from four complementary measurements, including the Internet edge and core.
First, we introduce a method to infer DNS amplification attacks at Internet eXchange Points (IXPs).  We exploit the central position of the IXP to
comprehend abused infrastructure dynamics %
and explore opportunities to fingerprint attack origins.
Second, we use a large, distributed honeypot platform to infer whether the IXP
and honeypots observe the same set of attacks, and to investigate if
attackers appear to exclude honeypots from attacks.
\autoref{fig:data_sketch} visualizes our extended perspective on inter-domain DNS amplification attacks.
Note that we anticipate attackers to abuse infrastructure that responds to DNS queries, which includes DNS~forwarders and recursive resolvers.
Third, we compare our observations 
with data from Internet-wide open resolver scans, allowing us
to assess the extent of existing views on abusable infrastructure.
Last, we consider comprehensive DNS measurement data to gain insights into the type of
DNS infrastructure abused (\ie open resolver versus authoritative nameserver) as well as the amplification potential of~attacks.

\smallskip
\noindent {\bf In detail, we address three key research~questions:}

\paragraph{Question 1 (\autoref{sec:disjoint_attacks})}
\textit{Does an IXP-centric view contribute additional insights into DNS amplification attacks?}

As we will show, passive observations of DNS-based reflection and amplification attacks at an IXP can identify
misused query names and abused infrastructure beyond
honeypot-based inferences.
Surprisingly, with an overlap of only $\sim4\%$, IXPs and honeypots detect
mostly disjoint sets of attacks.  In total, we find 24k new attacks over the
course of 3 months, which were not observed by the honeypots.

\paragraph{Question 2 (\autoref{sec:trace_attackers})}
\textit{Can we fingerprint outstanding attackers within the DNS amplification ecosystem?}

We fingerprint a larger attacking entity by correlating the use of
\texttt{.gov} names and static DNS transaction ID behavior.
The entity in question is demonstrably dominant and responsible for 59\% of inferred attacks.
Our data suggests two topological changes (\ie relocation) of the attacking
infrastructure within one~year, indicated by shifts in network layer observables.
We observe that the entity frequently changes abused
amplifiers. Moreover, we recognize patterns in misused query names that
strongly suggest attempts by the entity to improve the overall amplification
factors. %

\paragraph{Question 3 (\autoref{sec:vector_exploit})}
\textit{How efficient is the current exploitation of the DNS, meaning: (i) how are the amplifiers misused; and (ii) can the amplification factor still be improved?}

We are able to pinpoint the abuse of at least 10 to 1000~amplifiers in
most events.
Our results show that attackers mainly misuse legitimate \texttt{.gov} names in spoofed DNS queries, which is likely, because names under the \texttt{.gov} zone are DNSSEC
signed.
Bilateral clustering also shows that only 2\% of attacks use static amplifier lists.
95\% of the amplifiers for which we observe abuse are also found by a large-scale platform
that scans for abusable infrastructure, which suggests that attackers use mostly well-known, publicly
documented amplifiers.
Nevertheless, we reveal that 2\% of amplifiers are abused before they show up
in public scan data, suggesting that attackers also employ alternative methods to find amplifiers.

Overall, our observations show that attackers exploit amplifiers effectively,
and the turnover makes fine-grained source-IP filtering much harder. In
spoofed requests, attackers also misuse query names that lead to significant amplification
factors.  After inspecting 440~million domain names in DNS measurement data, we detect only 9000~names with
larger amplification potential.
At the same time, our estimation of DNS response sizes for these names reveals
that they could cause up to $14\times$ more amplification. This shows
that attackers do not fully exploit the DNS-based attack vector. %

In the remainder of this paper, we present background and related work in \autoref{sec:related_work}. We
outline four viewpoints from the complementary measurements in
\autoref{sec:data_sources}, and introduce our DNS attack detection method for an IXP in \autoref{sec:ixp_dns_attacks}.
We then proceed to answer our research questions in \autoref{sec:disjoint_attacks}--\autoref{sec:vector_exploit}, summarize discussions in \autoref{sec:discussion}, and conclude in~\autoref{sec:conclusion}.

\section{Background and Related Work}
\label{sec:related_work}

\paragraph{Reflective Amplification Attacks and Honeypots} Reflection and
amplification attacks~\cite{rowrs-adads-15} are
traditionally observed with honeypots~\cite{nwsks-shsda-16}, which apply straightforward thresholds
to infer attack activity and to discern mere scanning for
reflectors~\cite{kramer2015amppot, noroozian2016gets, thomas20171000}. The
advantage of using honeypots is that all incoming requests are likely part of
attacks or scans since legitimate DNS services do not send DNS queries to those sensors.
Honeypots, however, cannot infer the extent to which other infrastructures are
involved (\eg public DNS resolvers) and are therefore limited in the assessment of general attack properties such as intensity.
An additional challenge arises because the number of attacks visible to honeypots appears to converge quickly with only a few sensors deployed.
Deploying more sensors does not necessarily increase the breadth of
observation. This effect was shown with fewer than 10~sensors~\cite{kramer2015amppot}. Thomas~\emph{et al.}~\cite{thomas20171000} use a capture-recapture
approach to estimate a 85\%--97\% visibility into UDP reflection attacks. 

The research community has so far shown a tendency towards detection
techniques for edge networks~\cite{rossow2014amplification,
meitei2016detection}. We instead centre on IXP-based detection at the Internet
core. Only NTP-based attacks have been studied at IXPs~\cite{kopp2019ddos}
by explicitly launching attacks via an attacking infrastructure.  We
focus on attacks in the wild and on DNS-based reflection, which requires
a comprehensive detection mechanism. We also consider four Internet-scale,
complementary data sources to investigate attack visibility and attacker
behaviour.
Our approach allows us to refute the common assumption that (sizable) honeypot
infrastructures offer a near complete view on DNS-based reflection attack
activity.

Recent prior work~\cite{ddos2021kopp} started to compare attacks seen at an IXP and a honeypot using a flow- and volume-based DoS classification.
The authors found little overlap (8.18\%) between both vantage points.
Following up on this, we present the first in-depth comparison between various DDoS ecosystem viewpoints, precisely targeting DNS.

\paragraph{DNS Amplifier Ecosystem}
A DNS infrastructure that responds to \emph{all} incoming requests is prone to be
abused for reflection. This includes resolvers, forwarders,
and authoritative nameservers \cite{macfarland2015characterizing,anagnostopoulos2013dns,nksw-tfuco-21}.
DNS is the second most-common amplification vector, although its amplification potential
is $\sim$10$\times$ smaller compared to NTP~\cite{anagnostopoulos2013dns,
czyz2014taming} and it has the highest churn in reflectors among protocols
susceptible to reflection.
K\"uhrer~\emph{et al.}~\cite{kuhrer2015going,kuhrer2014exit} show that this is mainly caused by open resolvers in
access networks, \eg home routers, where dynamic address allocation leads to
the quick disappearance of about 50\% of identified amplifiers, when indexed by
IP address.

Anecdotal evidence suggests that attackers abandon reflectors once
response rate limiting (RRL) is detected \cite{winstead2014response}.  Due to
ample availability of DNS reflectors (2.1M in 2021~\cite{nksw-tfuco-21}), RRL can be
counteracted \cite{rossow2014amplification} by scaling. High reflector churn as
well as RRL force attackers to maintain and frequently update sizable amplifier
lists. This aligns with the observation that DNS exhibits the highest daily rate of
unique scanners \cite{thomas20171000}.
Exploiting our IXP-centric view, we follow the abuse of amplifiers over a
three-month measurement period, allowing us to unveil how efficiently attackers
deal with churn. Honeypot-based studies have to date not been able to
do so.

\paragraph{Forged DNS Queries and Names}
The query name and type in DNS queries affect the amplification factor.
Historically, the most common queries included unpremeditated as well as
crafted domain names, which were set up and used for amplification attacks
immediately after registration \cite{kramer2015amppot}.
\texttt{ANY} is an evident query type, yet querying for specific records can equally lead to large responses~\cite{kramer2015amppot, fachkha2014fingerprinting}.
DNSSEC is a DNS extension that enables verification of DNS content but
at the same time significantly increases the potential for amplification due to
larger response sizes~\cite{rvr2014dnssec,anagnostopoulos2013dns}.
Consequently, benign \texttt{.gov} names, which are subject to a DNSSEC
mandate~\cite{tossw-saawm-21}, started being misused in amplification attacks \cite{thomas20171000}.

We shed light on how attackers select names and study effective amplification
in attack traffic at the Internet core.  We also analyze large-scale DNS
measurement data to estimate the amplification potential of other names,
allowing us to reveal that while attackers are prudent in selecting names,
other choices would lead to higher amplification.

\paragraph{Origins of DNS-Based Attacks}
As reflection and amplification attacks involve IP~spoofing, attack attribution is
challenging~\cite{lichtblau2017spoofing,ehsw-rssdi-19,mlhcb-cisti-19,ocfmj-tdssi-20}.  In the case of NTP and its moderate amplifier churn, considering the
set of abused NTP servers has shown utility towards attributing attacks to a
DDoS-for-hire service~\cite{kopp2019ddos}. However, other research shows
that overlap in underlying infrastructure can exist, in addition to other obstacles
to fingerprinting~\cite{santanna2015booters}.
Not all honeypot sensors are necessarily used by attackers at the same time
and attackers can choose to abuse a subset of available reflectors in subsequent
attacks. Nevertheless, clustering methods such as KNN allowed researchers to
fingerprint a few major attacking entities and attribute attacks to
them~\cite{krupp2017linking, stress2016karami}.
Another commonly used feature is the IP Time-To-Live (TTL) field, which was
used to narrow down attack origins~\cite{backes2016feasibility,krupp2016identifying}.

Despite challenges in fingerprinting attackers, we successfully
use network and application layer data to fingerprint a major attack
entity, responsible for over half of the attacks detected at the IXP.
We follow this entity for over 9~months. %

\section{Complementary Data Sources}
\label{sec:data_sources}

We involve diverse and largely independent data from four data sources, bolstered
by orthogonal methods. We next provide an overview of our
main data to further comprehend the DNS amplification ecosystem,
Our starting point is data from an Internet
eXchange Point (IXP) for a three-month measurement period (2019-06-01 -- 2019-08-31). %

\subsection{Traces from a Large, Regional IXP}
\label{sec:data_ixp}

IXPs are a key component of the Internet to interconnect Autonomous
Systems (AS) without introducing high costs.
Observing traffic at a popular IXP provides a similar vantage point to that of large transit providers~\cite{sigcomm12_ixp}. 
We use traffic captures from a
large, regional
IXP in Europe.
Our IXP connects over a hundred member networks and observes 
traffic peaks of 600\,Gbps.
We now detail how we identify and sanitize DNS data in IXP traffic,
before using the data for attack detection.

\paragraph{Identifying DNS Traffic at IXPs} 
Our traces involve \texttt{1:16k} packet sampling and packets are truncated
after 128~bytes, which can be challenging with respect to analysis of
higher-layer (\eg ap\-pli\-ca\-tion-lev\-el) protocols. On the upside of things, DNS usually
operates with single UDP packets, hence packet sampling has no adverse effect
as we do not need to observe complete flows. Moreover, the first 128~bytes
of packets are sufficient to analyze DNS query packets.
On the downside, in terms of analyzing DNS answers, response data is usually
only partially visible (about 2 resource records per packet on average), since each DNS
response contains request as well as answer data.
Even though large UDP packets might be truncated and we cannot see the full answer data, we are still able to infer response packet sizes from the UDP length field, which precedes the DNS header.

Please note that we focus on DNS over UDP because TCP-based amplification
attacks do not exploit features of DNS but only inefficient implementations of
transport-layer sockets~\cite{kuhrer2014hell}.
Also, even though stubs and forwarders use more recent DNS variants (DNS over TCP/TLS/HTTP) to contact resolvers, a recursive resolver usually still uses UDP to reach authoritative nameservers.
TCP attacks use unencrypted traffic \cite{nxnsattack2021afek}.
During our measurement period, only 1.25\% of unencrypted DNS packets are based on TCP as a transport~layer.

\begin{figure*}
  \includegraphics[width=\textwidth]{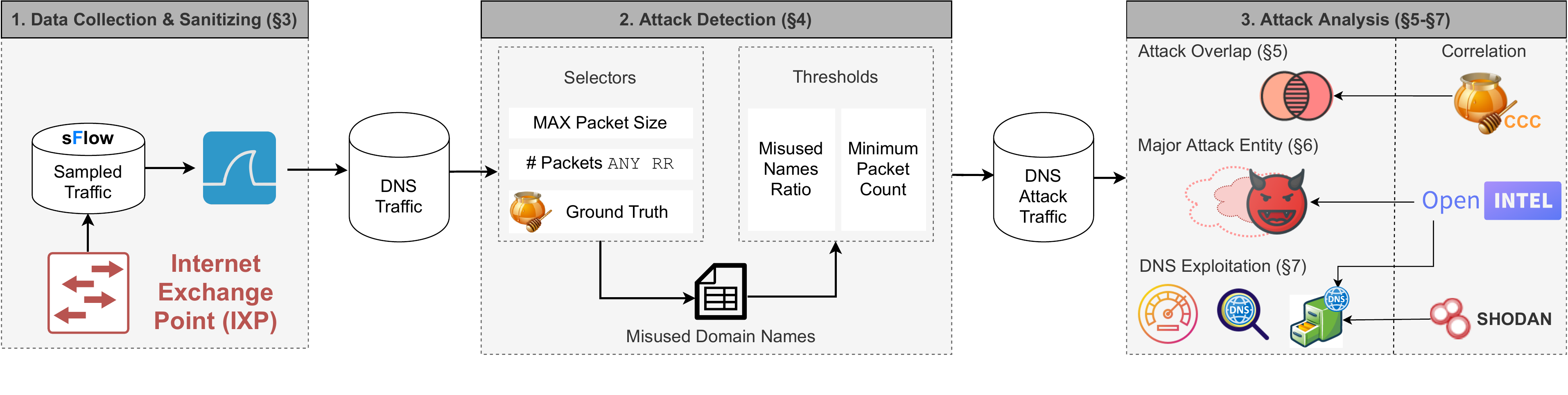}
  \caption{Overview of our data sources and attack inference steps.}
  \label{fig:toolchain-overview}
\end{figure*}

We use Tshark's DNS packet filter and dissector for protocol identification
and empirically verify that truncated DNS packets are identified correctly.
In the case of a UDP packet that leads to IP fragmentation, only the first
fragment is identified as it contains the DNS header.  This
effectively avoids double counting of fragmented DNS answers.
Overall, we find 33~million sampled DNS packets from June to September 2019, which correspond to a total of 528 billion DNS packets.

\paragraph{Sanitizing and Annotating DNS traffic} 
We only consider packets that include:
\one IP and UDP headers; and
\two well-formed values for IP addresses, UDP length, DNS query types and names, \ie values allowed and standardized by the respective RFCs.
In the process, we disregard 3\% of previously identified DNS traffic.
In the resulting data set, we observe slightly more requests than responses per
day (60\% are requests).  Daily aggregate packet counts follow a
weekly pattern with small changes during the weekends.  The most and second most
common DNS query types are \texttt{A}~(57\%) and \texttt{AAAA}~(13\%) records.
Using public routing data~\cite{RipeRisWebsite} and IXP member information we map the origin AS for
99\% of packets, and the peering hop AS for 96\%. 
For each query and answer packet, we also note the client and server IP addresses.

\subsection{Additional Data Sources}
\paragraph{Honeypot Data} %
We use data from the \emph{Cambridge Cybercrime Center (CCC)} honeypots~\cite{thomas20171000}, which are distributed and capture reflection attacks at the Internet edge.
This honeypot infrastructure has various features:
\one it provides topological diversity by using 80~active sensors that are distributed across 62~IP prefixes and 15~ASes; and
\two it emulates open DNS resolvers, which are responsive to reflection attempts, while not harming the Internet. %
We learn 31k DNS reflection attacks from the CCC data during the same 3 months.

It is worth noting that we carefully verified that the CCC~platform is able to make similar observations compared to related honeypot platforms (for details see \autoref{apx:honeypot-convergence}).

\paragraph{Large-scale, Active DNS Measurements}
To investigate to what extent attackers might achieve
amplification, we involve a longitudinal data source of daily DNS measurements
that accounts for names that are not necessarily misused in amplification attacks (yet).
We use data provided by the OpenINTEL project, which actively
measures about 65\% of the global DNS namespace, using well over
1200 zonefiles as a starting point~\cite{rjsp-hsila-16}. OpenINTEL queries
for a set of common resource record types, which allows us to map amplifier IP addresses to DNS infrastructure
and to estimate response sizes (\ie amplification factors). %

\paragraph{Internet-wide Scans}
To verify whether an end host provided DNS services in the past, we use data from the Shodan search engine~\cite{ShodanWebsite}.
These data include daily scans of the complete IPv4 address space to discover Internet services per IP address.

\begin{table}%
\centering
\caption{Our various data sources backed by complementing methods to analyze DNS amplification attacks.}
\label{tab:data_sources}
\begin{tabular}{lcr}
\toprule
Data Source 		& Type	& Viewpoint\\
\midrule
IXP			& Traffic 	& Transit, Internet core \\
CCC Honeypot	& Traffic 	& Amplifier, edge network \\
OpenINTEL	& Scans 	& DNS TLD zone walking \\
Shodan		& Scans 	& Complete IP address space \\
\bottomrule
\end{tabular}

\end{table}

We summarize our data sources in \autoref{tab:data_sources}. Data sources of the category \emph{scans}
are based on active measurement methodologies, whereas
\emph{traffic} is brought about by passive observations.  The complementary
viewpoints allow us an in-depth understanding of effects observed for the
DNS amplification ecosystem, as we show in the following sections.

\section{Inferring DNS Amplification Attacks at an IXP}
\label{sec:ixp_dns_attacks}

We first introduce our methods to infer misused DNS names and DNS amplification attacks in IXP traffic traces. %
We then briefly report about using these methods for live monitoring.
\autoref{fig:toolchain-overview} shows an overview of our processing steps.

\begin{table*}
\caption{Distribution of attacks and attack traffic for misused names. \texttt{.gov} names that dominate amplified DNS traffic.}
\label{tab:tld_atk_count}
\centering
\begin{tabular}{llllllllllllll}
\toprule
TLD &    \texttt{.gov} &    \texttt{.za} &    \texttt{.cc} &    \texttt{.pl} &    \texttt{.cz} &   \texttt{.com} &   \texttt{.org} &    \texttt{.se} &    \texttt{.eu} &    \texttt{.be} & \texttt{root(.)} &    \texttt{.br} &     \texttt{.ru} \\
\midrule
\# Names   &     17 &     1 &     1 &     1 &     1 &     2 &     2 &     1 &     1 &     1 &      1 &     1 &      2 \\
\% Packets &  74.92 &  1.32 &  3.92 &   1.1 &  1.17 &  1.31 &  0.99 &  0.54 &  0.38 &  6.23 &   6.73 &  1.38 &  0.005 \\
\# Attacks &  22758 &  3969 &  3863 &  3732 &  3712 &  3388 &  3316 &  2663 &  2385 &  1551 &   1120 &   184 &      2 \\
Max. Size [B]  &   8069 &  5155 &  4408 &  5954 &  5881 &  10270&  6090 &  5535 &  4096 &  8199 &   4098 &  3893 &    -- \\
\bottomrule
\end{tabular}
\end{table*}

\subsection{Identifying Misused Names}
In DNS reflection attacks, queries for the right combinations of domain names and resource
record types can trigger large responses and hence lead to sizable
amplification. For this reason, attackers are likely to use effective names
recurringly. Based on this assumption we find a list of suspicious DNS query names.

We develop the list of names using three so-called selectors.  Two of our
selectors consider features in the IXP data. The third selector involves the
CCC honeypot data. The CCC data accounts for a substantial number of reflection attacks
for which we may observe attack traffic at the IXP.

\paragraph{Selector 1: Max Packet Size}
Our first selector considers the maximum (response) packet size of each and
every query name observed at the IXP. Note that the response size per name may
vary over time and also depends on the query type.
We rank query names such that the first selector can pick, \eg the top-ten names in
terms of max packet size.
Large DNS responses may lead to IP fragmentation. 
Even in the presence of fragmentation, the UDP header, however, allows us to
determine the size of the DNS response (see \autoref{sec:data_ixp}).  
The largest response of more than 10k bytes was triggered by an \texttt{RRSIG} query, the remaining top-ten largest responses are triggered by \texttt{ANY} queries. 

\paragraph{Selector 2: Number of \texttt{ANY} Packets}
The \texttt{ANY} query type is a convenient way to bring about DNS amplification,
provided that \texttt{ANY} queries are not restricted by the authoritative nameserver of
the chosen query name.  This is why our second selector considers names that
most appear in \texttt{ANY} query packets.
The ten top-ranked names according to Selector 2 are used almost exclusively for
\texttt{ANY} queries. %
Considering \texttt{A},
\texttt{AAAA}, and \texttt{ANY} packets, the share of type \texttt{ANY} packets is higher than $>$99.99\%
for all names but for the root (\texttt{.}) name ($97\%$).

\paragraph{Selector 3: Query Names Used Against Reflected DDoS Victims}
For our third selector we start by extracting all DNS attack victim IP
addresses and timestamps from the CCC sensor data. Next, we search for the IXP
DNS traffic associated with the attacks. Selector 3 then chooses the most
common names used in the traffic in question. We find DNS attack traffic for
16\% of all CCC DNS attack events ($\approx$~4.4k~victim IP addresses). We
identify two reasons for invisible CCC attack traffic at the IXP:
  \one The traffic is not routed via the IXP,
  and \two the traffic is routed via the IXP but the packets are not sampled (given our \texttt{1:16k} sampling rate).
The ground truth attack traffic consists almost exlusively of \texttt{ANY} packets (99\%) and we observe only 482~unique names with this selector.

We consider the IXP DNS traffic associated with victim IP addresses at the time
of an attack as \emph{ground truth}. We later use this ground truth
to validate detection thresholds. It is worth noting that the CCC data also provides
query names, however, we decide to not use them in favour of selecting names
that are actually visible from the perspective of the IXP in ground truth
attacks.

\begin{figure}%
  \begin{center}
  \includegraphics[width=0.99\columnwidth]{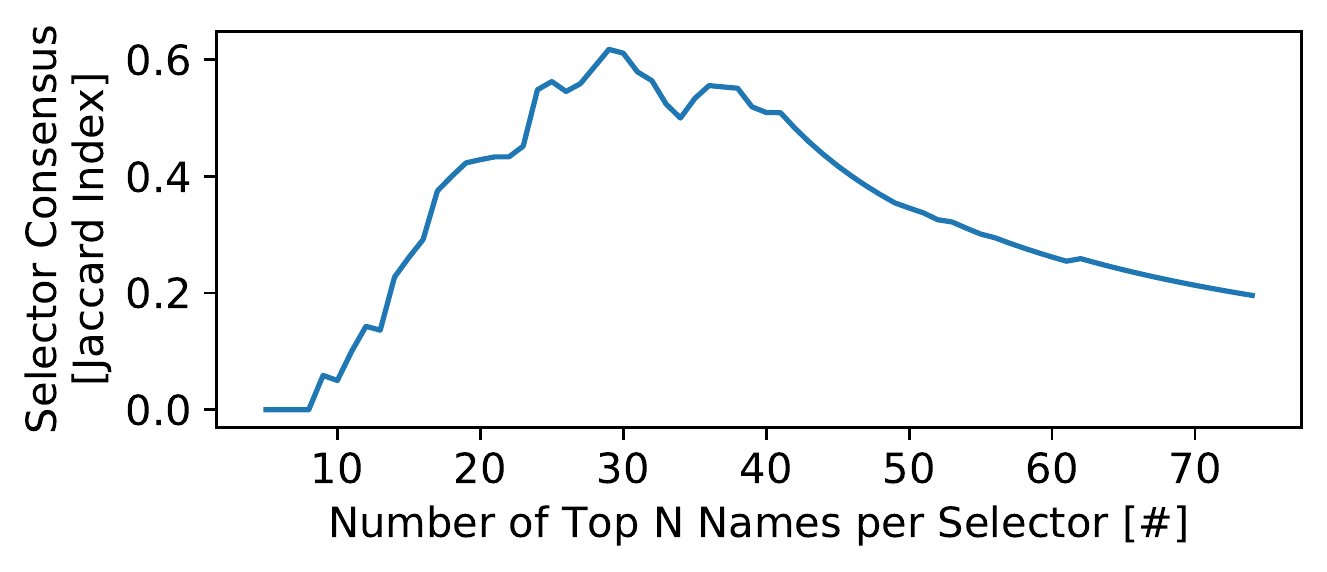}
  \caption{Our selectors detect the same names with a different ordering up to a set size of 29~names. These names are most likely to be misused in attacks.}
  \label{fig:metric_jaccard}
  \end{center}
\end{figure}

\paragraph{Number of Names per Selector}
The number of names chosen per selector is configurable. To determine the highest
similarity, we calculate the Jaccard index for the three sets of names using
increasing set sizes.
We observe a high consensus for 29~names per selector (see
\autoref{fig:metric_jaccard}), which shows that the first 29~names chosen by
each selector are almost the same, but with a different ordering.
Note that selecting the point of highest consensus is a conservative measure
for two reasons. First, this reduces the number of distinct names but chooses
names for which up to three selectors agree. And second, the selector
results follow a long-tail distribution with the knee points before the
consensus point, which means that selecting more names would lead to adding
\emph{insignificant} names.
All things considered, we set the size to detect misuse at 29~names per
selector.

Finally, we merge the three selector sets of names to create our
final list of names. The union combined with the high consensus point allows
for a conservative name selection while still keeping significant names detectable
only by a single selector. Our final list contains 34 names.
For 32 of these names (94\%), we detect attack traffic
(see \autoref{sec:atk_det_with_names}),
which demonstrates the effectiveness of the selectors in identifying misused names.
\autoref{tab:tld_atk_count} shows properties of the considered names, most
of which are part of the \texttt{.gov} zone.
21 names are mutually detected by all 3 selectors.
The intersection of Selector 1 and 3 contains three names, and the intersection of Selector 2 and 3 contains five names.
We find two exclusive names with Selector 1.
Overall, the IXP and the first two selectors are sufficient to create our list. %
The CCC data does not add any names compared to the unions and intersections of all three sets.
Using the honeypot-based selector is a good verification of the first two selectors, though, since it is based on ground truth
attack~traffic.

\begin{figure}[t]
  \begin{center}
  \includegraphics[width=0.99\columnwidth]{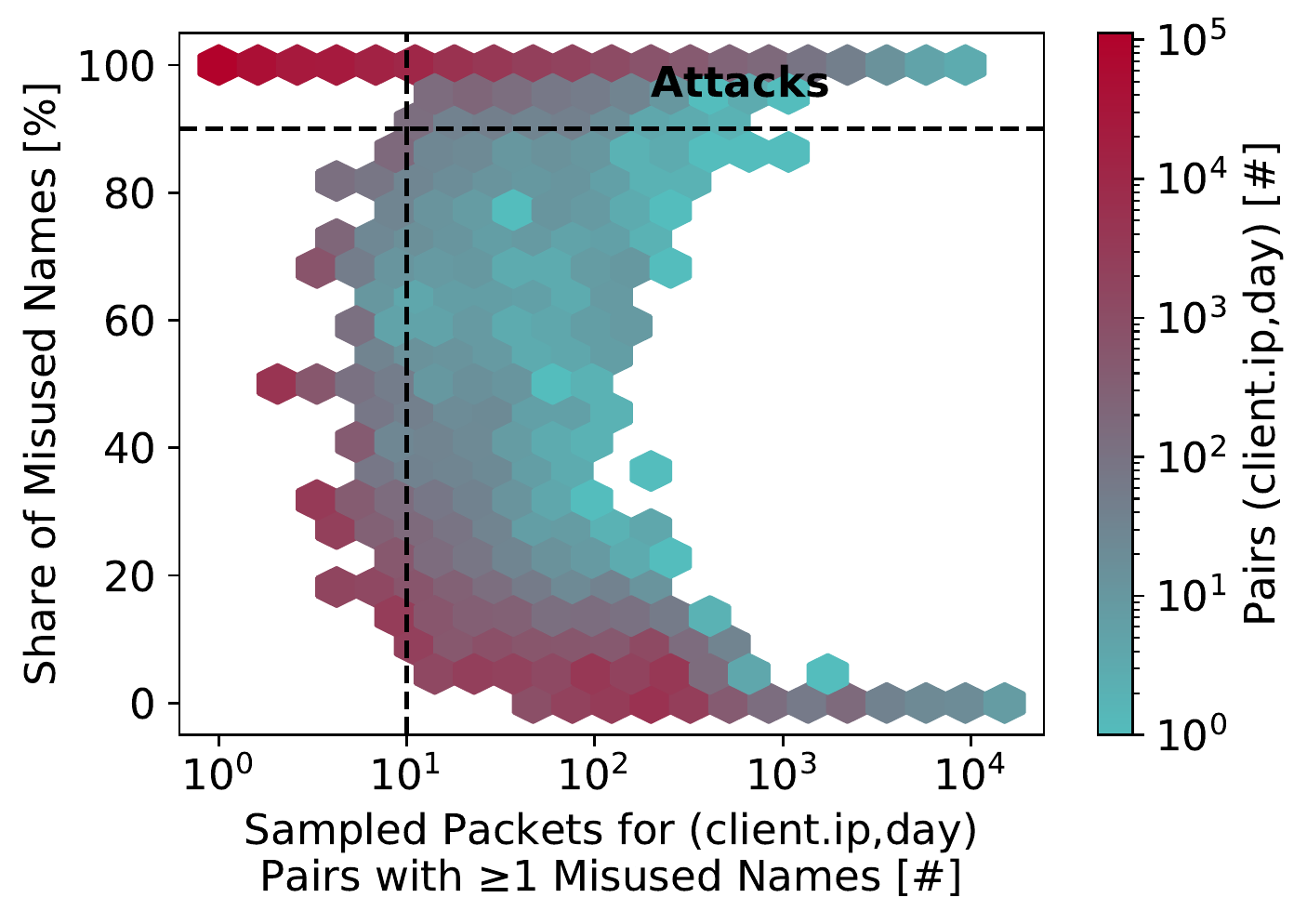}
  \caption{Share of misused names compared to overall traffic. Many clients exchange DNS traffic with only misused names, which aids attack detection.}
  \label{fig:ips_share_blacklisted_names}
  \end{center}
\end{figure}

\subsection{Attack Detection with Misused Names}
\label{sec:atk_det_with_names}

Using the previously introduced list of misused names, we now further analyze DNS packets that contain
queries or answer data for these names. This will allow us \one to define two thresholds for the detection of attacks at the IXP, and \two to group related packets into attacks.

\paragraph{Threshold 1: Traffic Share of Misused Names}
We calculate the daily ``traffic share'' of suspicious packets for client IP addresses (\ie supposed DNS query originators).
A high share can indicate attack activity.
Please note that the client IP address denotes the source IP address of DNS requests and the destination IP address of DNS responses.
The share of suspicious packets is calculated for each unique \textit{(client.ip,day)} pair for which at least 1 suspicious packet was observed.
This excludes unrelated DNS activity, \ie clients which exchange traffic for only benign names on a given day.
Then, we visualize the share of misused names for each \textit{(client.ip,day)} in \autoref{fig:ips_share_blacklisted_names}.
This reveals that with an increasing packet count a bimodal distribution becomes more pronounced,
\ie even though clients exchange large numbers of DNS traffic, the related traffic consists of only misused names or almost none.
The low shares occur due to the fact that one of the misused names is the root (\texttt{.}) name, which is also a very common name for legitimate DNS traffic.
This distinctive distribution allows for the introduction of thresholds to
detect attacks. With our first threshold, we define that a client is under
attack if the share of misused names exceeds 90\% on a given day. This finds
extreme cases of suspicious traffic shares but still
allows for a small error margin, \ie we might observe other names due to
legitimate DNS traffic of the client. Note that for clients with a low traffic
volume (\eg 1 sampled packet), this single threshold is not enough since it
most-likely leads to many false positives. We therefore set a minimum packet
count threshold at the beginning of the bi-modal distribution (details see Threshold~2). With respect to the minimum packet
count threshold, the traffic share threshold of 90\% accounts for the smallest
possible error, \ie exactly 1 sampled legitimate packet.

\begin{figure}[t]
  \begin{center}
  \includegraphics[width=0.99\columnwidth]{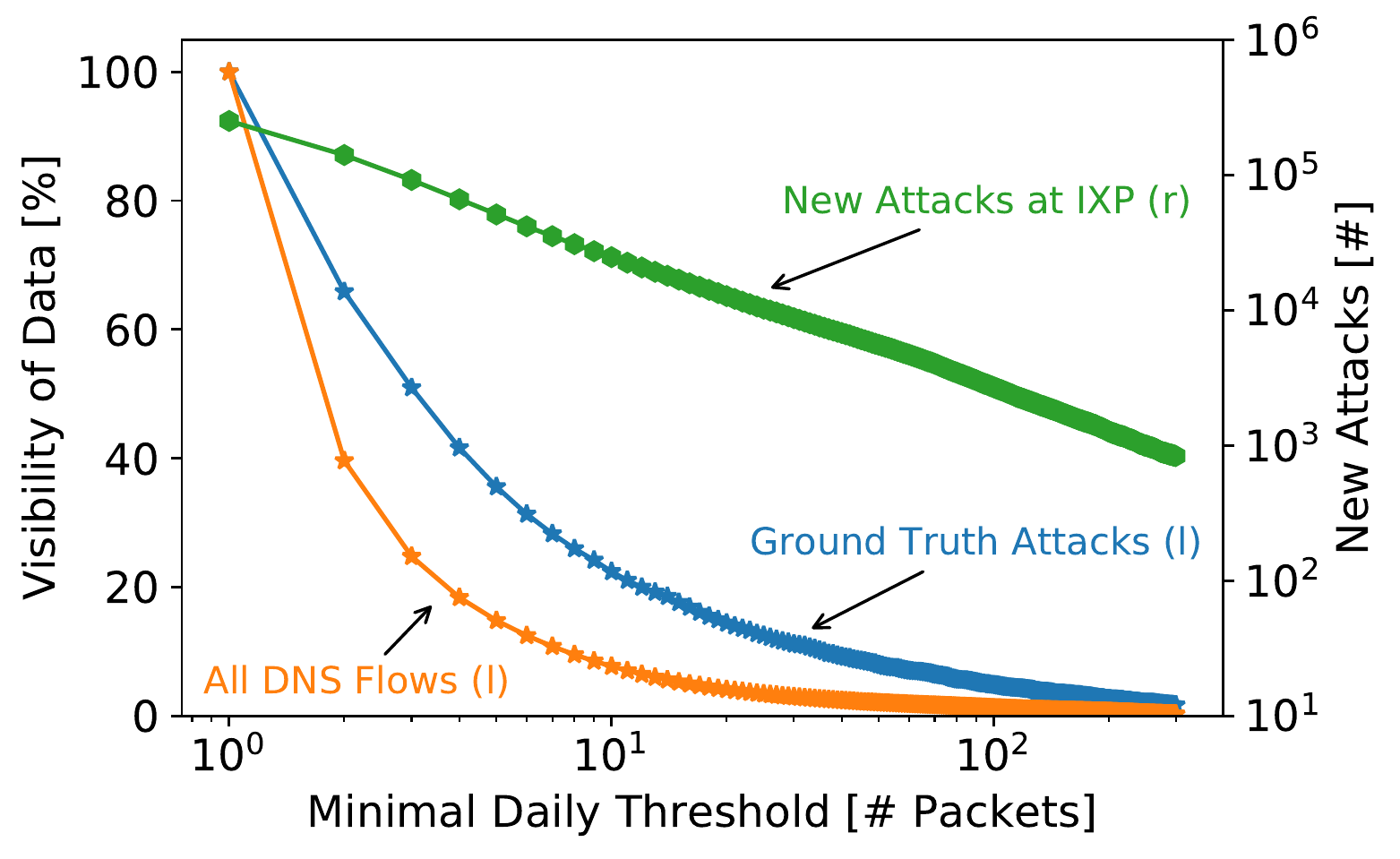}
  \caption{Visibility of all DNS flows and ground truth attack flows depending on the number of packets considered. Number of detected DNS attacks at the IXP based on the thresholds are shown on the right $y$-axis. }
  \label{fig:dns_visibility}
  \end{center}
\end{figure}

We argue that the high traffic share of misused names is a strong indicator of
attack traffic.
To illustrate this argument consider ten sampled
packets, our sampling rate of \texttt{1:16k}, and a misused name share of 90\%.
This would correspond to 144k packets with only misused names. No
client should reasonably exchange so much DNS traffic for legitimate reasons,
especially in the presence of DNS caches.

\paragraph{Threshold 2: Minimum Packet Threshold}
We now explore the effects of a minimum packet threshold at the IXP, in
particular we analyze the trade-off between the detection of all attacks
(\emph{visibility}) and reducing false positives. To this end, we use our
ground truth attack events that we found at the IXP with the help of CCC sensor data. We
count the number of packets for these attack events and plot the fraction of
visible events w.r.t. minimal packet count, see \autoref{fig:dns_visibility}.
To provide a reference point, we also include the visibility of DNS traffic for all \textit{(client.ip,day)} pairs.
Overall, this plot demonstrates which share of DNS traffic remains visible at the IXP if a minimum packet threshold is applied.
We find that 22\% of visible ground truth attacks exhibit at least 10 sampled packets,
\ie they remain detectable while applying a minimum packet threshold of 10~packets. Note that for all
\textit{(client.ip,day)} pairs, the visibility for 10 packets is, as expected,
much lower (8\%), since regular DNS flows only consist of significantly fewer
packets. Looking at the total number of additionally detected attacks at the
IXP (secondary $y$-axis), the threshold of 10 packets at minimum strongly limits
the number of detected attacks. We argue that this significantly reduces false
positives (or at least vague cases) but still allows us to find over 24k new,
significant attack events at the IXP. Again, these attacks were not observed by
the honeypots, hence provide an opportunity for new insights.

\begin{figure}[t]
  \begin{center}
  \includegraphics[width=0.99\columnwidth]{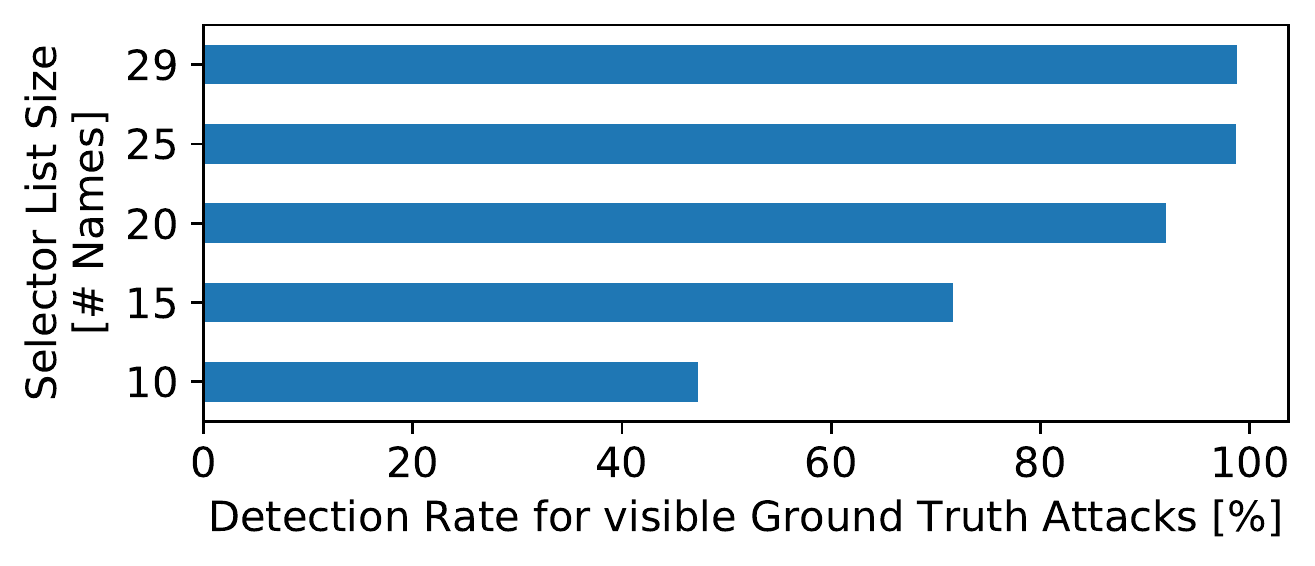}
  \caption{Attack detection rate based on selector list sizes and 2 thresholds. We reach 99\% for visible ground truth attacks and see a convergence around 29 names.}
  \label{fig:blacklist_size_attack_detection}
  \end{center}
\end{figure}

\paragraph{Validation}
A sound attack detection mechanism should be able to detect attacks for which we know to be visible in the sampled IXP traces. 
Given this notion, we now investigate the detection rate for visible CCC attacks, based on the defined thresholds and a varying number of names for our selectors.  
This allows us to verify whether precision of our attack detection would increase by adding more misused names. 
\autoref{fig:blacklist_size_attack_detection} shows that the detection rate converges at 99\% with 29 names per selector for our threshold configuration. 
This clearly illustrates that we do not need to fine tune our detection method further.
Also, this result is coherent with the selector consensus, which, again,
suggests that adding more misused names does not have a beneficial~effect.

\paragraph{First Glimpse into Detected Attacks}
At the IXP, we found 25.7k~attacks to
19k unique client IP addresses, which includes 24.6k new attacks (as previously
mentioned).
The detected attacks are dominated by traffic created by misusing \texttt{.gov} names,
see \autoref{tab:tld_atk_count}. The attack durations match the observations of
security reports~\cite{NetscoutSecurityReport, KasperskySecurityReport} with
many short-lived attacks (25\% shorter than 7 minutes 50\% shorter than 33
minutes).
One third (36\%) of the total attack traffic is sent towards victims in ISP
networks, which is the largest victim group after content networks (24\%).

We see no signs of NXNS attacks \cite{nxnsattack2021afek}.
Those rely on responses including NS referrals with many NS names (>30) but no glue records.
In our data, 70\% of the responses include at most 1 NS entry and 90\% at most 10 NS entries.
Recently detected attack vectors (\ie \texttt{SRV}, \texttt{URI}), which also offer a $10\times$~amplification factor \cite{ampmap2021moon}, are also not used, yet.

\subsection{Live Monitoring}

We deployed our method at the IXP to verify online detection capabilities in realistic settings.
Our prototype consists of two building blocks:
\one A module that identifies potentially misused names in near real-time.
\two A module that continuously analyzes changes compared to the previous day.
Without advanced performance optimizations, we are able to identify misused names within a maximum delay of 5~minutes, on commodity hardware.

We utilize our deployment to assess victim and name fluctuations.
Overall, we see quite stable numbers of unique victims and also very stable
lists of misused names. %
On average, we observe 631~unique victim /24-prefixes (492 /16-prefixes and 121 /8-prefixes) per day.
The name lists have a mean Jaccard index of 0.96 in comparison to the respective previous day.
This suggests that daily updates for misused names are not necessary; we keep them to identify changes quickly.

\section{\mbox{Comparing~IXP~and~Honeypot~Data}}
\label{sec:disjoint_attacks}

We now present basic properties from attacks inferred at the IXP and
compare the observations to honeypots.
We find that the IXP and the honeypot sensors observe a vastly disjoint
set of attacks.  Both vantage points share only 1.1k attack events, which
corresponds to 4.2\% of all events at the IXP and to 3.5\% of 31k attack events
at the honeypots.
This is a surprising result, given that prior work~\cite{thomas20171000,kramer2015amppot} assumed that a distributed honeypot, such as ours, can capture a large percentage of global reflection attacks.
We consider an IXP~vantage point to be an opportunity to observe DNS amplification attacks which
have been so far invisible to the research community, and potentially provide
new insights, \eg for attacks that deliberately exclude honeypot platforms.%

\begin{figure}[t]
  \begin{center}
  \includegraphics[width=0.99\columnwidth]{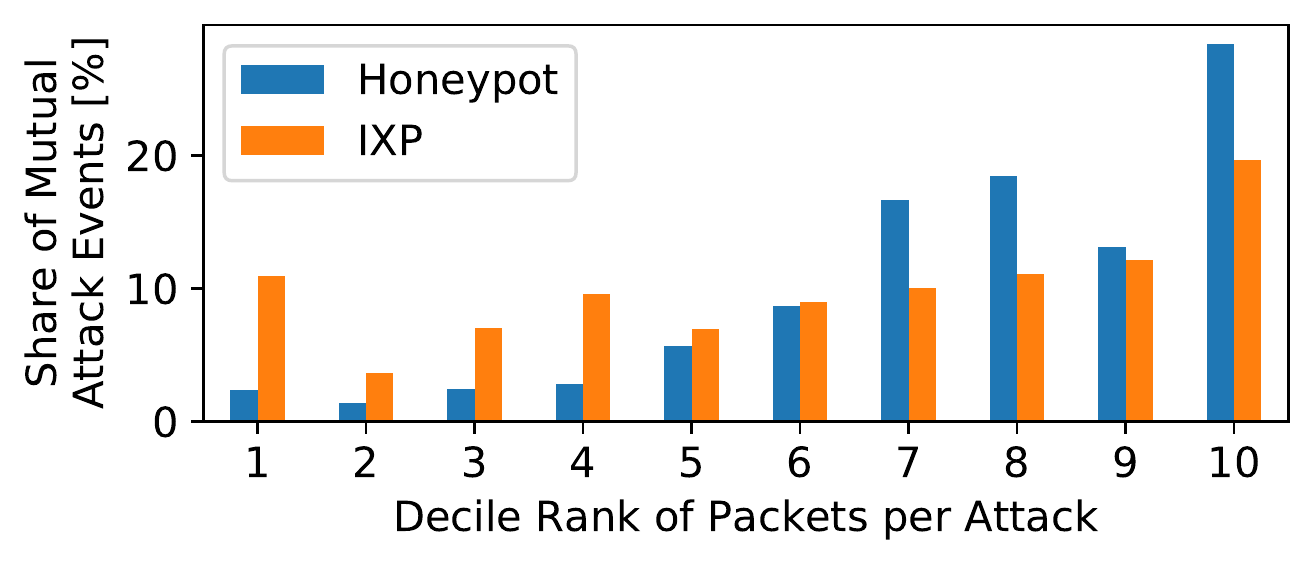} %
  \caption{Attacks detected by the IXP and honeypots (1098) differ in relative attack intensity score: Mutual attacks are rather strong honeypot attacks, but medium-sized IXP attacks.}
  \label{fig:atk_intensity_mutual}
  \end{center}
\end{figure}

While the overlap is small, we now check, for comparative purposes, whether the
honeypots and the IXP agree on the observations for mutual attacks.  To this
end, we calculate a relative attack intensity score for all attacks that have
been identified by each type of vantage point.  We do this by sorting all attacks by the total
packet count and calculating the deciles.  Then, we rank each attack with the
decile score of 1 to 10.
We plot the relative distribution of intensities for mutual attacks in
\autoref{fig:atk_intensity_mutual}.  Overall, honeypots are rather sensitive
vantage points: the mutual attacks are mostly strong honeypot attacks (with a
mean intensity of 7.7) and medium-sized IXP attacks (with a mean intensity of
6.3).
We argue that this is due to packet sampling and our thresholds, which make
smaller attacks invisible at an IXP.  Hence, honeypots are good vantage points
to detect small-sized attacks, \emph{if} they are abused by the attacker as
reflector. 
IXPs, on the other hand, show that a substantial number of large attacks occur, which were not observed by the honeypots.
They are likely to see even more small attacks but this would require significantly smaller sampling, which is uncommon in practice. We will leave this for future work.

\section{Tracing a Major Attack Entity}
\label{sec:trace_attackers}

In this section, we reveal a major attacking entity, which is responsible for 59\% of all attacks at the IXP.
To this end, we identify recurring patterns based on the selection of domain names, the creation of DNS requests, and the selection of amplifiers.
Our method does not depend on our specific vantage point but can be generalized to other inter-domain data sets.

We link multiple independent events to an attack entity.
We explicitly use the abstract term \emph{entity} as we do not refer to a specific botnet, booter website \etc but to the essence that maintains an infrastructure to select names and amplifiers to launch attacks.

\subsection{Fingerprinting Using Domain Names}
\label{sec:major-attack-name-fingerprinting}

\begin{figure}[t]
  \begin{center}
  \subfigure[Attack traffic volume of misused names based on IXP data.]{\includegraphics[width=0.49\textwidth]{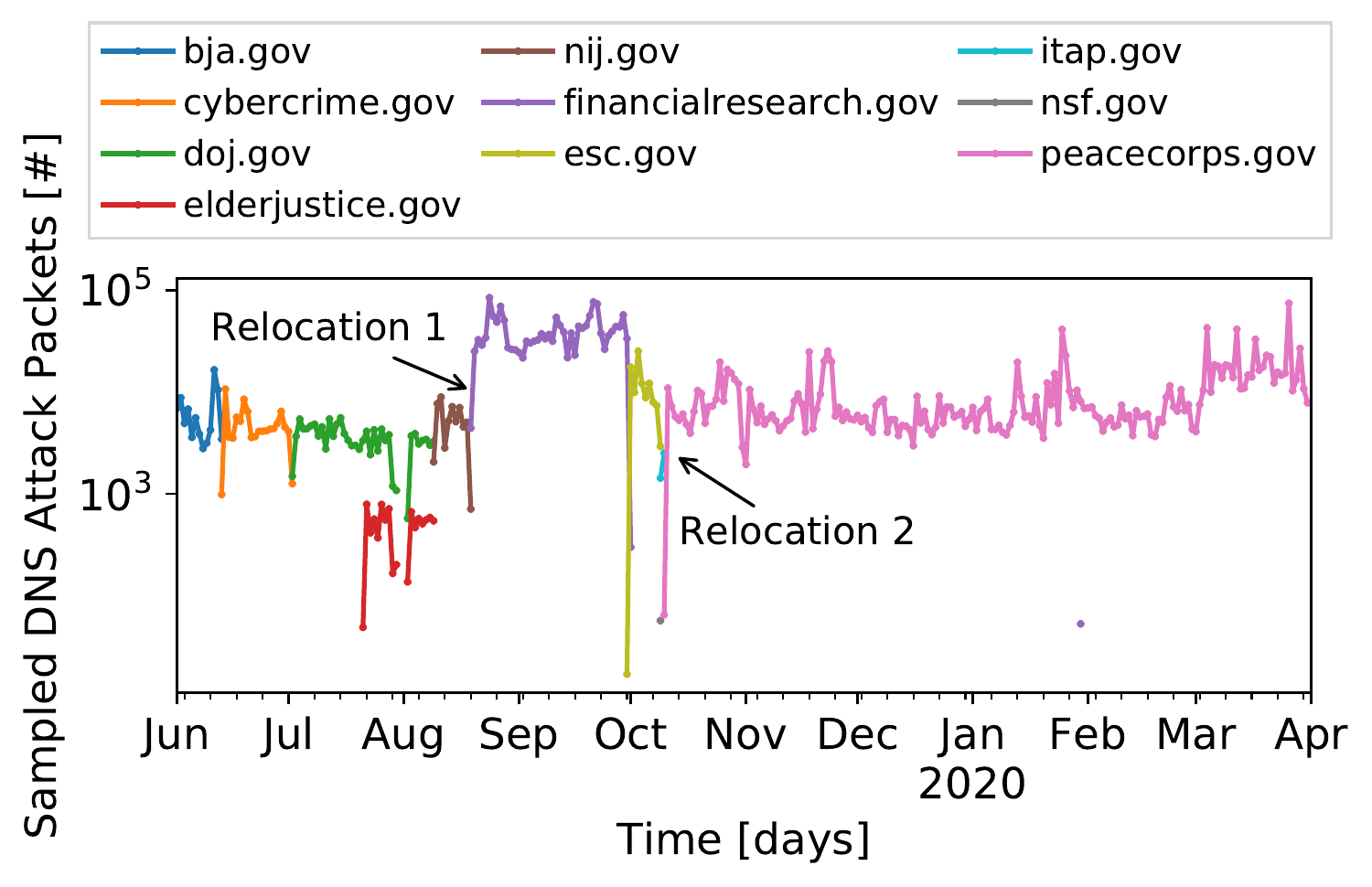}\label{fig:ts_entity_names_complete}}
  \subfigure[Estimated \texttt{ANY} response sizes of currently misused names based on OpenIntel data set. Plateaus indicate DNSSEC key rollovers. Cutoff of a single anomaly (value of 12.5kB) for better readability.]{\includegraphics[width=0.49\textwidth]{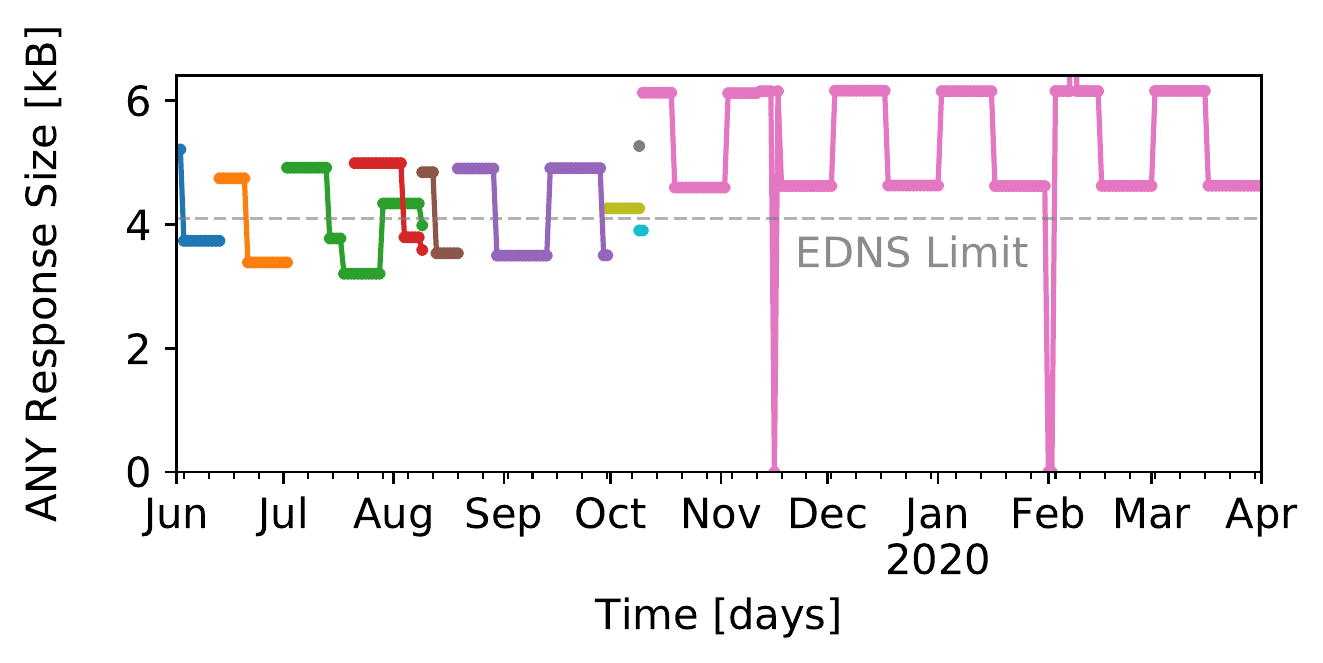}\label{fig:openintel_any_size_plateaus}}
  \caption{Time series of synchronized names misused by major attack entity.}
  \label{fig:ts-names}
  \end{center}
\end{figure}

We conduct a time series analysis of the misuse of names.
Our results reveal a clear transition between names for selected \texttt{.gov} names (see \autoref{fig:ts_entity_names_complete}), which contribute 59\% of the overall attack traffic.
Names appear to be chosen in lexicographical order, except for few weeks in which two names were used concurrently.

The transition pattern between names strongly suggests that a specific
entity is involved in the attacks. Independent misuse of the
same domain would not lead to clear, abrupt transitions.
Interestingly, we observe an increase in attack traffic at the IXP following changes in misused name.
This hints at a driving factor behind the transitions in lexicographical order.
To further investigate this observation, we analyze the expected response sizes offered
by names, as this is crucial for amplification attacks.
\autoref{fig:openintel_any_size_plateaus} depicts the \texttt{ANY} response
sizes of each name inferred from the OpenINTEL data set, which provides us with
historical DNS data.  The dashed line indicates the
recommended maximum payload size (4096~bytes) of EDNS~\cite{RFC-6891}, the
extension mechanism in DNS to carry, \eg DNSSEC data.
{\bf We observe that the expected response sizes change while names are
actively misused in attacks, and also that transitions to other names follow drops
in sizes.}
Further analysis of the OpenINTEL data set reveals that the plateaus in response sizes---which last two weeks---relate to DNSSEC key rollovers.
When a new zone signing key (ZSK) is introduced, an increase in response size can be expected, as multiple \texttt{DNSKEY} records are present at the same time.
ZSK rollovers can be completely automated in software, which explains the regular patterns.
RFC 6781 \cite{RFC-6781} recommends two rollover schemes, pre-publish and double-signature. 
Pre-publish introduces only the new key in stand-by mode, \ie the key is not yet used to sign RRsets.
This allows resolvers to learn about the new key before it is actively used.
This scheme, however, is prone to race-conditions and misconfigurations \cite{chung2017dnssec} which impair the validation process.
To overcome the challenges of pre-publishing, the double-signature scheme has been introduced.
Double-signature allows two active ZSKs and generates two (redundant) \texttt{RRSIG} records signatures.
The old ZSK can then be retired at any given time.
On the downside, this scheme doubles the number of signatures in a zone. %
Although both rollover schemes are proposed in RFC~6781 \cite{RFC-6781}, pre-publish has been established as a de-facto standard.
It was used \texttt{4x} more often than double-signature in 2016 \cite{chung2017dnssec, pft2021dnssec}, and \texttt{8x} more often in 2020~\cite{pft2021dnssec}.
Also, it is recommended by various DNS software vendors \cite{pft2021dnssec}.

We only observe double-signature schemes for the misused \texttt{.gov} names.
This leads us to conclude that {\bf operators of these \texttt{.gov} names, many of which are US federal government domain names, do not only not adhere to best practices, which exacerbates amplification, but also that these decisions introduce misuse by others.}
A recent (Q2~2021) sample of DNSSEC records for \texttt{nsf.gov} and \texttt{doj.gov} shows that the rollover practices did not~change.

Even though we observe transitions after a (variable) number of days when the expected size is below the recommended EDNS limit (see the valleys in \autoref{fig:ts-names}), we cannot reasonably infer the decision making process behind.
Either the attack entity completely understands DNSSEC mechanics or simply observes $<4096$~byte responses and then (manually) transitions to the next name.

\begin{figure}%
  \begin{center}
  \includegraphics[width=0.99\columnwidth]{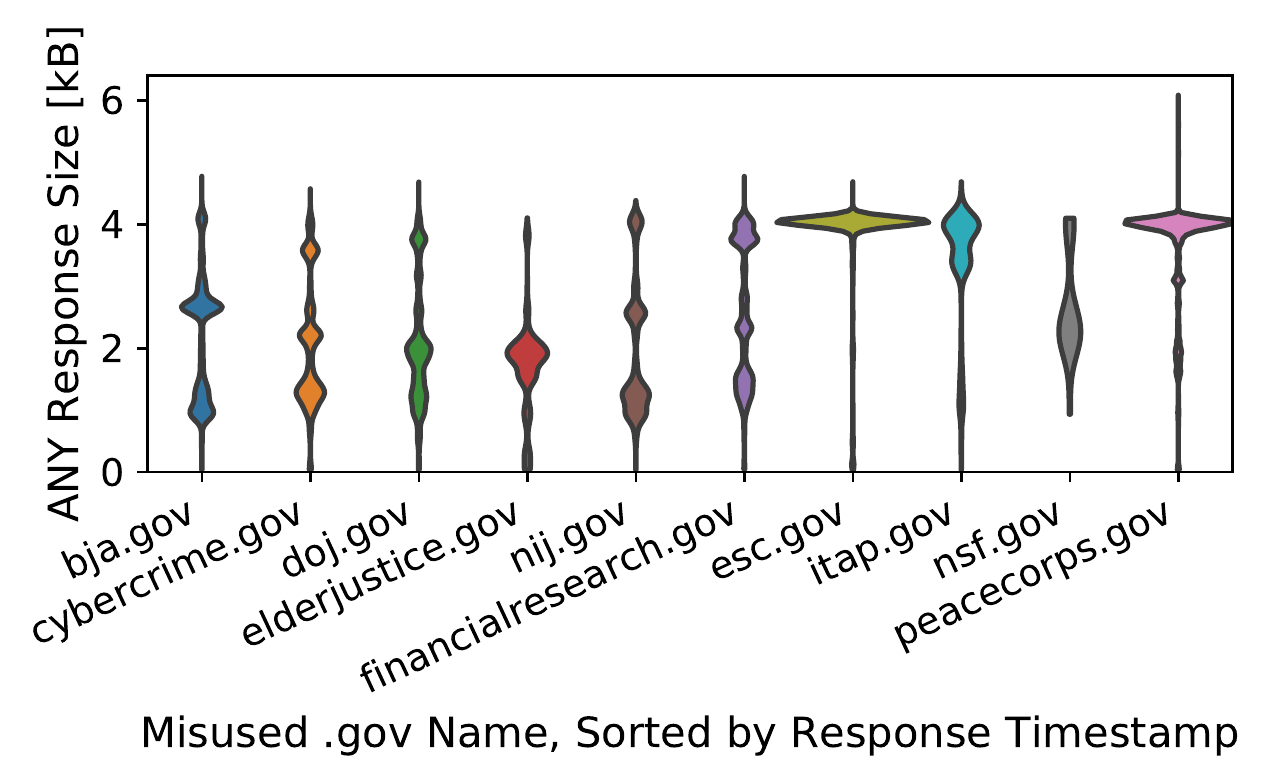}
  \caption{Violin plot of the observed DNS response sizes at the IXP for the major attack entity.}
  \label{fig:violinplot_gov_names_main_followup_vertical}
  \end{center}
\end{figure}

By analyzing the packet sizes in the sampled IXP data, we can confirm that the attack entity achieves effective amplification factors.
In contrast to \autoref{fig:openintel_any_size_plateaus}, which exhibits the potential maximum \texttt{ANY} response sizes, \autoref{fig:violinplot_gov_names_main_followup_vertical} shows the relative frequency of the actual response sizes observed at the IXP and extracted from the UDP headers, grouped by each name.
Please note that we consider all DNS~query types for the misused names.
In the attack traffic, however, we only observe the type \texttt{ANY} for these names.
Most names exhibit a bi- or tri-modal distribution.
The observed clusters of response sizes near the theoretical limit highlight that the attack entity succeeds in finding names (and related authoritative nameservers) as well as amplifiers that still allow \texttt{ANY} requests.
Closer investigation reveals that smaller response sizes appear rather at the end of a name's life cycle.
This bolsters our result that the entity observes the current effective amplification factor and updates misused names upon a decline.

\paragraph{Additional Fingerprinting Features}
To verify that we can link multiple events to a single entity, we seek other features that may indicate uncommon consistency over time.
To this end, we perform an entropy analysis of packet header fields that usually should show high randomness.
If they do not, we suppose the deployment of the same attack tool.
Pre-built headers and usage of raw sockets may lead to such consistent behavior, for example.

For each attack event, we check whether the number of distinct values of a specific feature
grows linearly with the total packet count. 
We investigate header fields such as \texttt{IP.ID}, \texttt{UDP.SRCPORT}, and \texttt{DNS.ID}.
As network and transport layer features change after reflection, we consider only DNS queries (\ie packets that are sent before amplification) here.

\begin{figure}[t]
  \begin{center}
  \includegraphics[width=0.99\columnwidth]{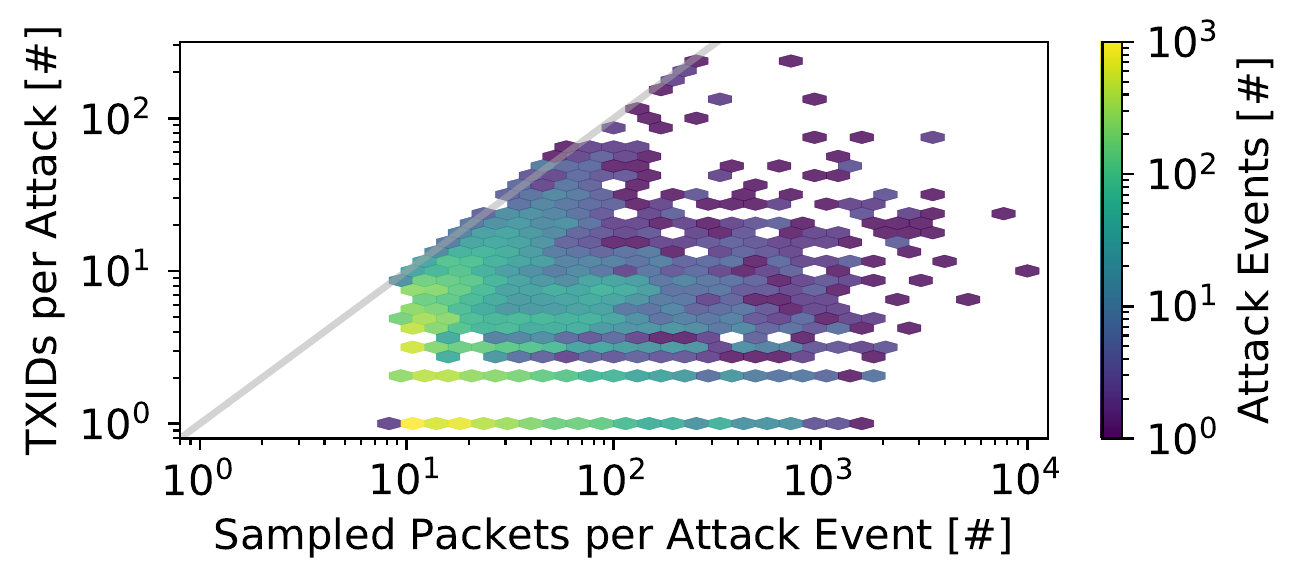}
  \caption{Entropy check: \# of unique DNS transaction IDs and packets for all packets from attack entity. A limited number of DNS IDs indicates pre-built queries.}
  \label{fig:booter_scatter_dns_all_flows}
  \end{center}
\end{figure}

Unfortunately, all features in the network
and transport layer headers exhibit a linear growth and hence a high
entropy.  Fortunately, we detect a low randomness for the DNS transaction ID, a feature in the application header. The number of IDs in use is
usually 1-2 orders of magnitude smaller than the total packet count, see
\autoref{fig:booter_scatter_dns_all_flows}.
The low entropy gives good reasons to manually inspect the DNS~IDs. 
We found that 91\% of attack events
have only a (seemingly random) selection of odd or even IDs. With respect to
the minimal number of sampled packets containing misused names ($9$), the probability
for this observation with random DNS IDs is $2\cdot(1/2)^9 = 0.4\%$. Also, we rule
out measurement artifacts such as a synchronization between traffic and
sampling, since sampling selects \texttt{1 out of 16k} and not \texttt{every
16kth} packet.
Hence, we argue that we found an arithmetic structure
and not only a random phenomenon. For the remaining 9\% of attack events we
observe two phases with odd and even IDs, respectively, and a distinct shift.
Indeed, the overall selection of IDs for the attack events
with synchronized names follows a two-day rhythm, alternating between odd and
even DNS transaction IDs every 48 hours, independent of other features.
The selection of IDs is probably seeded with timestamps and not linked to the properties of the victim. 

In summary, we are able to fingerprint a major attack entity based on two
properties: \one its selection of names and \two the
implementation details of the attack tool (selection of DNS IDs). Both features are part of the application header, which means that we can
link attack traffic to this entity even after the reflection~occurred.
Similar to our observations in \autoref{sec:disjoint_attacks}, the fingerprint of this attack entity is only visible for $\le 0.6\%$ of the attacks detected by the honeypot,
\ie the attack entity is only clearly visible at the~IXP.

\subsection{Attacked Victims, Misused Amplifiers}
\begin{figure}[b]
  \center
  \includegraphics[width=0.99\columnwidth]{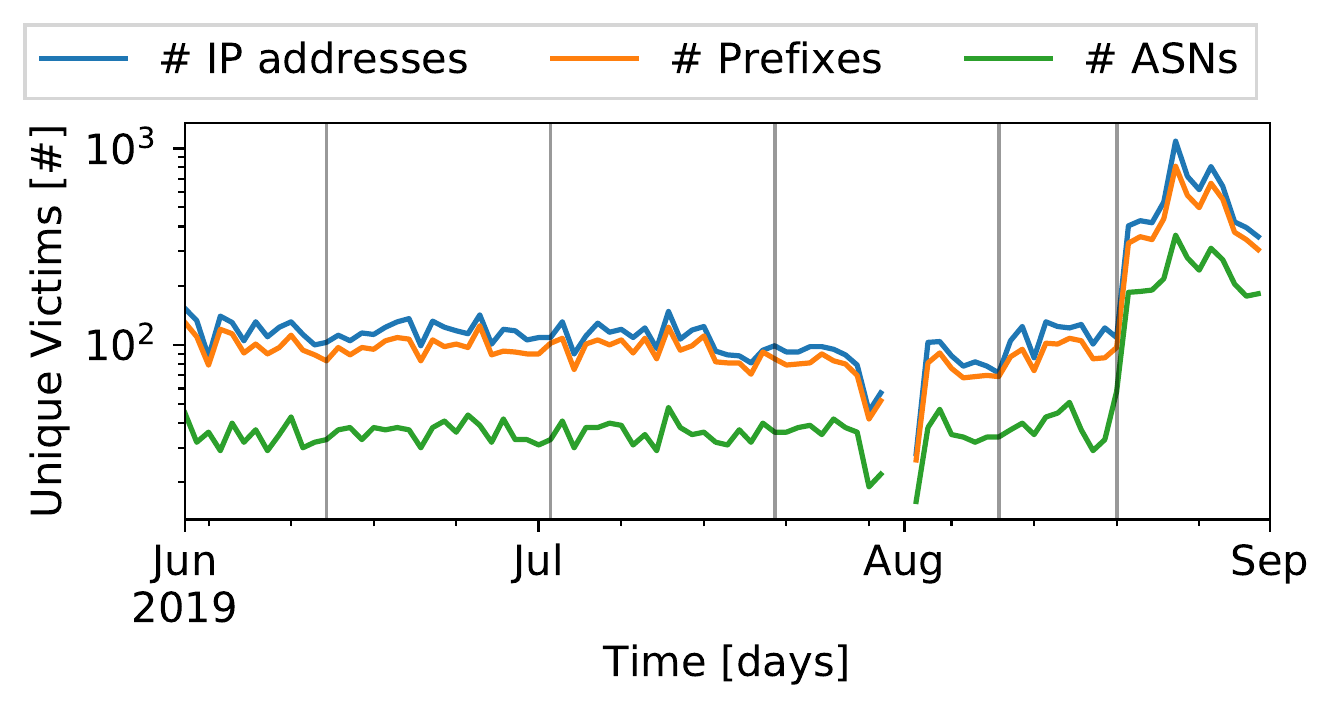}
  \caption{Number of unique victims identified by IP address, prefix, and AS numbers. Vertical lines highlight transitions of misused names.}
  \label{fig:ts_booter_victims}
\end{figure}

We now describe the executed attack events as well as reconfigurations and relocations controlled by the attacker.
The attacks we associate with this entity are distributed. We observe almost
as many victim destination IP addresses as covering victim prefixes per day,
see \autoref{fig:ts_booter_victims}. In this plot, we highlight the transitions
between misused names with vertical lines. The number of victims remains 
stable until the transition to the last name occurs.
Then, the number of victims increases by almost an order of magnitude.
The increase also correlates with the total number of packets (compare \autoref{fig:ts_entity_names_complete}).

\begin{figure}[t]
  \begin{center}
  \includegraphics[width=0.99\columnwidth]{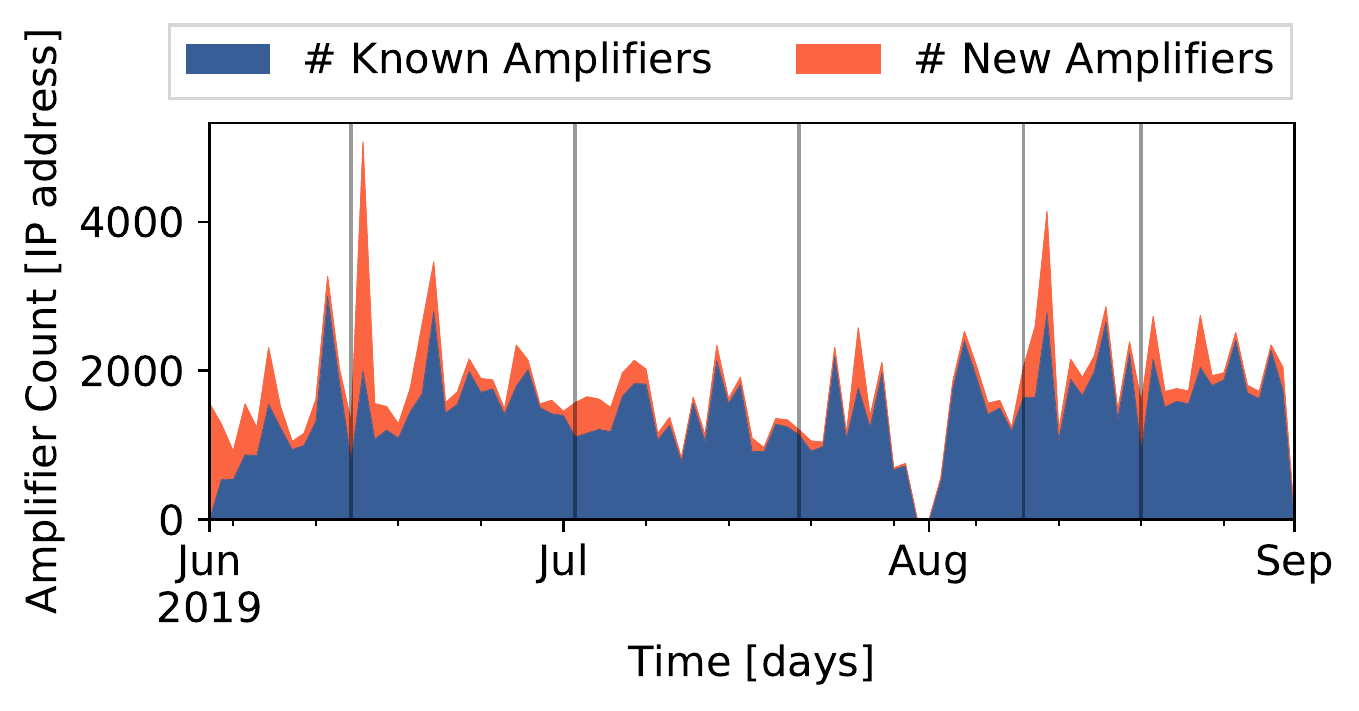}
  \caption{Known and new amplifiers used by the major attack entity.
      Bursts of new amplifiers correlate loosely with name transitions
      (vertical~lines). 
      }
  \label{fig:new_amplifiers_single_origin}
  \end{center}
\end{figure}

We check whether the attacker reconfigures only the list of misused names or also the list of misused
amplifiers. To this end, we count the daily number of new and known amplifiers,
\ie amplifiers that have been already misused at least
once, see \autoref{fig:new_amplifiers_single_origin}. This plot introduces two
findings. First, although the number of total attacks increases, the number of
misused amplifiers remains stable. This suggests that the entity misuses only a
specific set of amplifiers per day. Random subsets, however, are selected per
attack event, which we will show in detail in \autoref{sec:vector_exploit}. Second, periods with significantly more new amplifiers usually
follow name transitions, indicating that names as well as the amplifier list
were updated at the same time. Nevertheless, new amplifiers appear almost
daily, revealing a more continuous update behaviour, which is necessary due to
IP~churn of DNS~hosts.

In order to understand the increased number of attacks, we continue
investigating other features. We find that starting with the peak mid of August, the DNS
request-response ratio for this entity shifts dramatically. Before, we observe
almost purely amplified DNS traffic, \ie DNS responses.
The absolute numbers of responses remain stable, however, we see a stark increase in requests.
Now, $\sim$~85\% of attack traffic consists of requests.
Moreover, 99.8\% of the requests
originate from the same ingress AS and exhibit the same IP TTL of 250. Seeing
such a concentration from a single ingress AS indicates a centralized attack
infrastructure, because botnets are usually distributed across multiple networks.
 Such infrastructures are usually the hidden back-end of
booter websites. Unfortunately, the customer cone of this ingress AS contains
more than 16k ASes, so we are not able to fully trace back the infrastructure.
We do not find topological changes at the IXP that would justify the shift in
attack traffic properties, \ie we do not find any new members and the ingress
network has been already a member during the whole measurement period. Also, it is
unlikely that this shift is caused by unrelated routing updates as the paths to
all amplifiers would have to change simultaneously for such a homogeneous
effect. Instead we argue that since the shift occurred concurrently with a name
transition, this effect is triggered by the attack entity itself, namely due to a
relocation into the ingress cone of the IXP member.
We define relocation as the topological transition of a centralized attack infrastructure into another network.
We later observe a second relocation in mid October.

To sum up, by revisiting the main measurement period with a fingerprint in hand, we were able
to identify reconfigurations of names and amplifiers, and a relocation of the attacking
infrastructure. We were able to do this on the basis of network and application
layer information visible at an IXP.

\begin{figure}[b]
  \center
  \subfigure[Number of amplifiers per attack: Most victims are attacked by 10 to 100 amplifiers.]{\includegraphics[width=0.95\columnwidth]{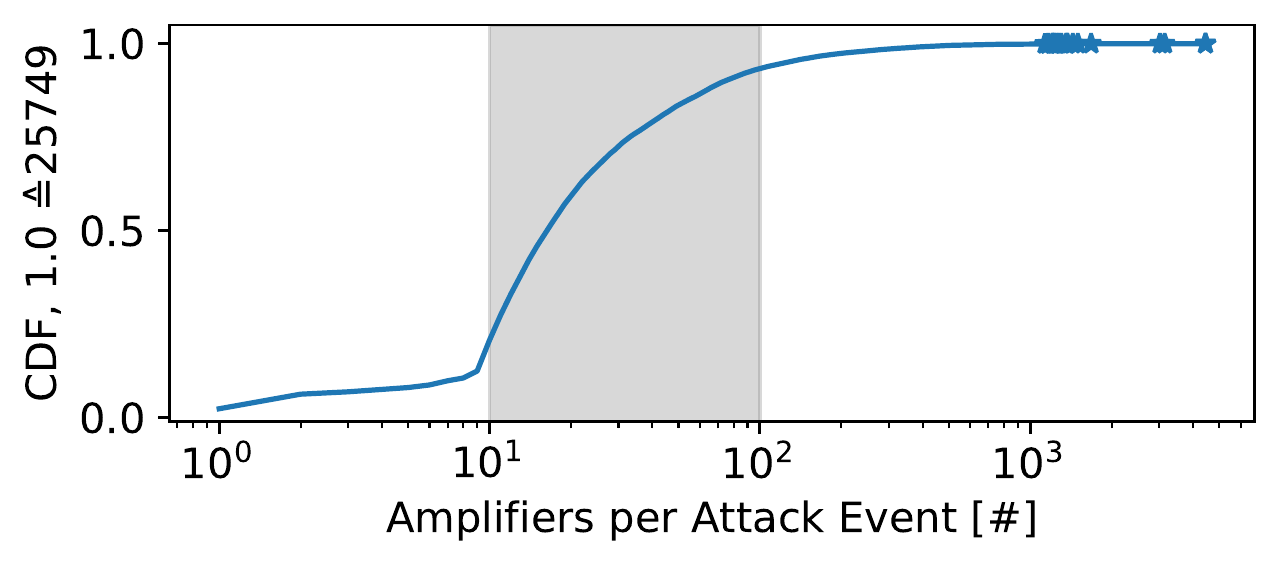} \label{fig:cdf_num_reflectors_per_victim}}
  \quad
  \subfigure[Number of attacks per amplifier: 50\% of the amplifiers participate in more than one attack, even 23\% of amplifiers in more than ten attacks.]{\includegraphics[width=0.95\columnwidth]{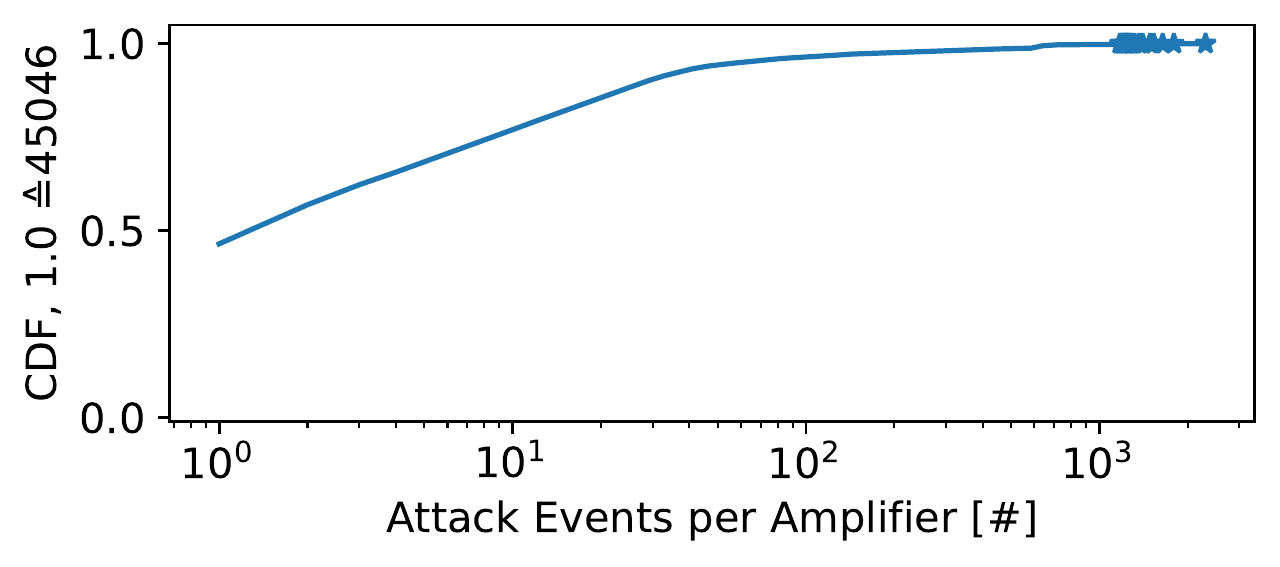} \label{fig:cdf_num_attacks_per_reflector}}
  \caption{Distributions of amplifier involvement in attacks. Last 20~data points are highlighted.}
  \label{fig:amplifiers-attack-events}
\end{figure}

\section{Unveiling DNS Attack Practice}
\label{sec:vector_exploit}

In our last analysis, we use our inter-domain IXP perspective to disclose the abused infrastructure.
Also, we analyze how (all) attackers perform in terms of amplification~efficiency.

\subsection{Amplification Ecosystem}

We start by investigating whether attackers continue to abuse the same reflectors across attacks. Repeatedly using a stable 
infrastructure may make attackers \one fingerprintable  and \two susceptible to frequent re-addressing of edge resolvers.%

\paragraphNoDot{How many amplifiers are used in attacks? And in how many attacks do particular amplifiers appear?}
The OpenINTEL data accounts for a large number of authoritative nameservers
active during our main measurement period: approximately 4.2 million
\texttt{NS} names that together map to well over a million IP addresses.
We use these data to associate  amplifier IP addresses observed at the IXP with
authoritative nameservers, where applicable.

We find that only 908 authoritative nameservers are abused in attacks---about 2\% of all amplifiers observed at the IXP. 
By exclusion, we conclude that the vast majority of abused DNS amplifiers are open resolvers or forwarders. 
We discuss further a classification of forwarders and resolvers in \autoref{apx:cache-snooping}.
This does not come as a surprise, because authoritative servers should not recursively 
resolve DNS queries, which makes them less attractive reflectors.
Root-query-based attacks, however, utilize 4$\times$ more authoritative nameservers, which can be linked to attacks misusing misconfigured root hint-files~\cite{DNSRootVariationAttacks}, \cite[Chapter 4]{aitchison2011pro}.
We observe that 80\% of attack events use between 10 and 100~amplifiers (numbers not extrapolated by sampling rate), \emph{cf.}
\autoref{fig:cdf_num_reflectors_per_victim}.
Also, \autoref{fig:cdf_num_attacks_per_reflector} shows 23\% of amplifiers that participate in more than ten attacks. Such recurrent use of
amplifiers may allow for fingerprinting attackers.

\paragraphNoDot{Do attack entities work with stable lists of amplifiers?}
After observing recurrent amplifiers, we now investigate whether 
 attackers use relatively static amplifier sets. As the DNS is subject
to high amplifier churn~\cite{kuhrer2014exit} from home gateways with 24h IP address
lease times~\cite{kuhrer2015going}, we expect sets to exist for short time spans, only.

We approach our analysis by quantifying the (dis)similarity of two attacks from measuring the
Jaccard distance over its respective sets of amplifiers.  A group of similar
attacks (\ie cluster) with a low Jaccard distance among each other indicates a
fixed list.  We use a bilateral clustering method by using two well-known
algorithms: T-Distributed Stochastic Neighbor Embedding (T-SNE)~\cite{maaten2008visualizing} and
Density-Based Spatial Clustering of Applications with Noise (DBSCAN)~\cite{ester1996density}.
We compute both algorithms independently to exclude a biased result from a single
clustering method.

\begin{figure}[t]
  \begin{center}
  \includegraphics[width=0.99\columnwidth]{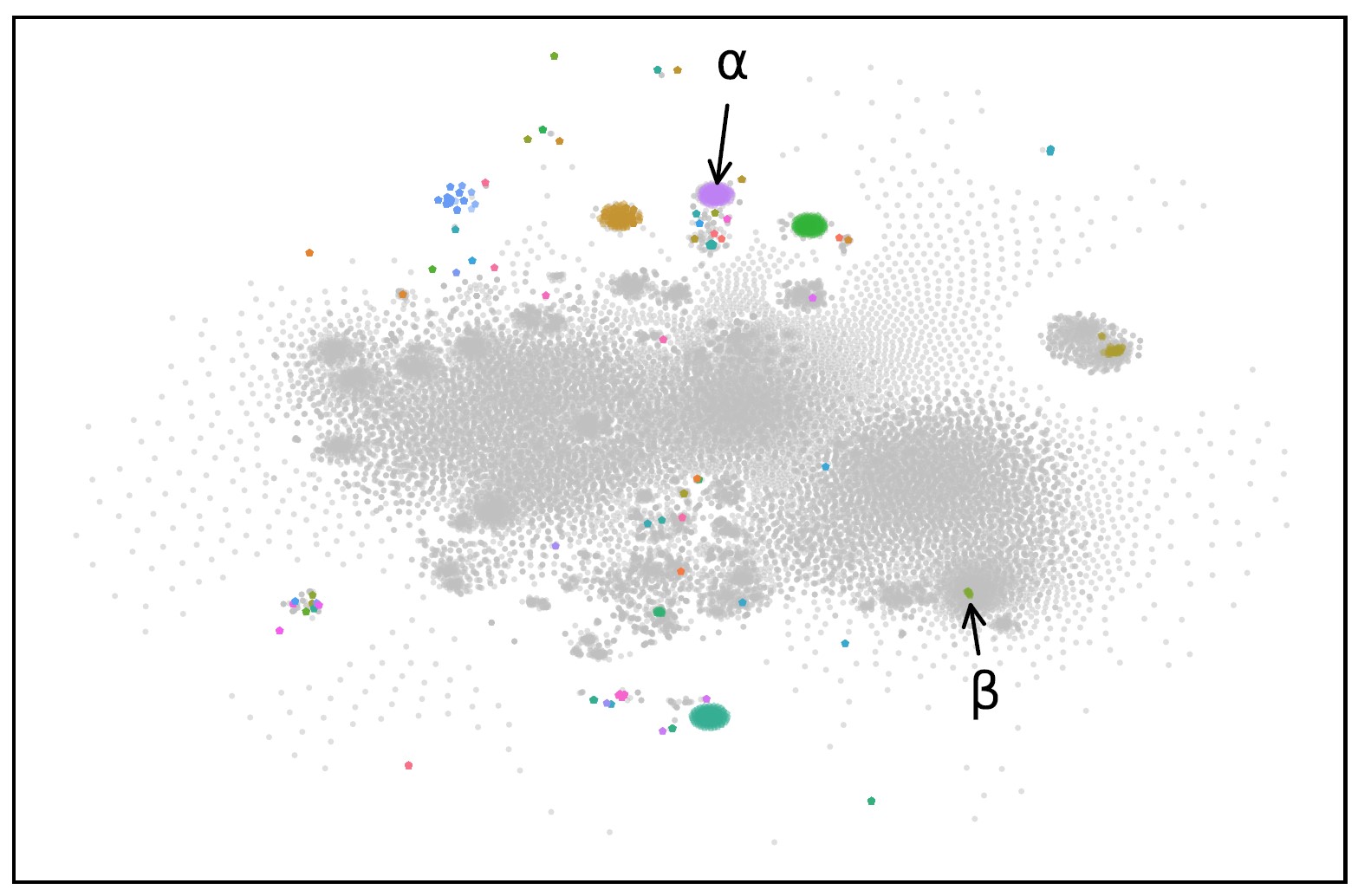}
  \caption{T-SNE visualization of attack events based on Jaccard distance over the amplifier sets. DBSCAN clusters marked with colors (gray being not classifiable). Both clustering algorithms agree on the dissimilarity of attack events.}
  \label{fig:clustering_tsne_dbscan}
  \end{center}
\end{figure}

T-SNE allows us to visualize high-dimensional data on a two-dimensional plane.
Similar attacks are moved towards each other and dissimilar
attacks are move apart.  
We observe very stable results for different 
perplexity parameters.
The clustering results are visualized in \autoref{fig:clustering_tsne_dbscan},
each gray scatter point represents a single attack event. T-SNE indicates a
strong dissimilarity between most events, with some noticeable clusters.  DBSCAN
groups nearby neighbors into clusters and marks non-classifiable outliers
within low-density regions. 
  We next combine both
clustering results in the single \autoref{fig:clustering_tsne_dbscan}.
DBSCAN cluster IDs are encoded with colors, the non-classifiable outliers in
gray.

We see 67 clusters while $\sim$92\% of attack events remain outliers. 
We inspect clusters of at least 5 attacks and 5 amplifiers to find
stable sets. Here, the most static amplifier set ($\alpha$) was used for 177~attacks during 40 days without any change. The largest set ($\beta$)
uses $\sim$527 amplifiers per attack while always introducing a small,
steady change.  We can attribute in total only 2\% of attack events
to fixed sets.  Attackers seem to steadily use a random combination of known
and new amplifiers.  This reinforces our previous findings that attackers leverage the
amplification ecosystem and that source-based filtering is infeasible to mitigate DNS amplification attacks.

\paragraphNoDot{Do attack entities recruit new amplifiers?}
Since our results suggest that attackers steadily vary 
their amplifier sets, we question which amplifiers are used over time.
To this end, we use the Shodans historic lookup, which allows to retrieve
its complete scan history for a given IP address. Shodan omits transparent DNS forwarders.
It lists currently around 2 million recursive DNS resolvers, which all can be abused for reflection.
Next, we perform a historic IP address lookup for all 45k~amplifiers
observed at the IXP.

We find that 95\% of these amplifiers are reported by Shodan to serve recursive DNS at some point in time.
This finding grants two insights:
\one it confirms independent observations that although most amplifiers are known to the community, we fail to remove these amplifiers ultimately~\cite{park2019you,nksw-tfuco-21}; 
\two attackers do not use private but mostly publicly indexed amplifiers.
Scan results can differ even for Internet-wide measurements, \eg due to the origin of the scanner~\cite{wan2020origin}.
However, both Shodan and attackers observe a very similar set of DNS-amplifiers.

\begin{figure}[t]
  \center
  \includegraphics[width=0.99\columnwidth]{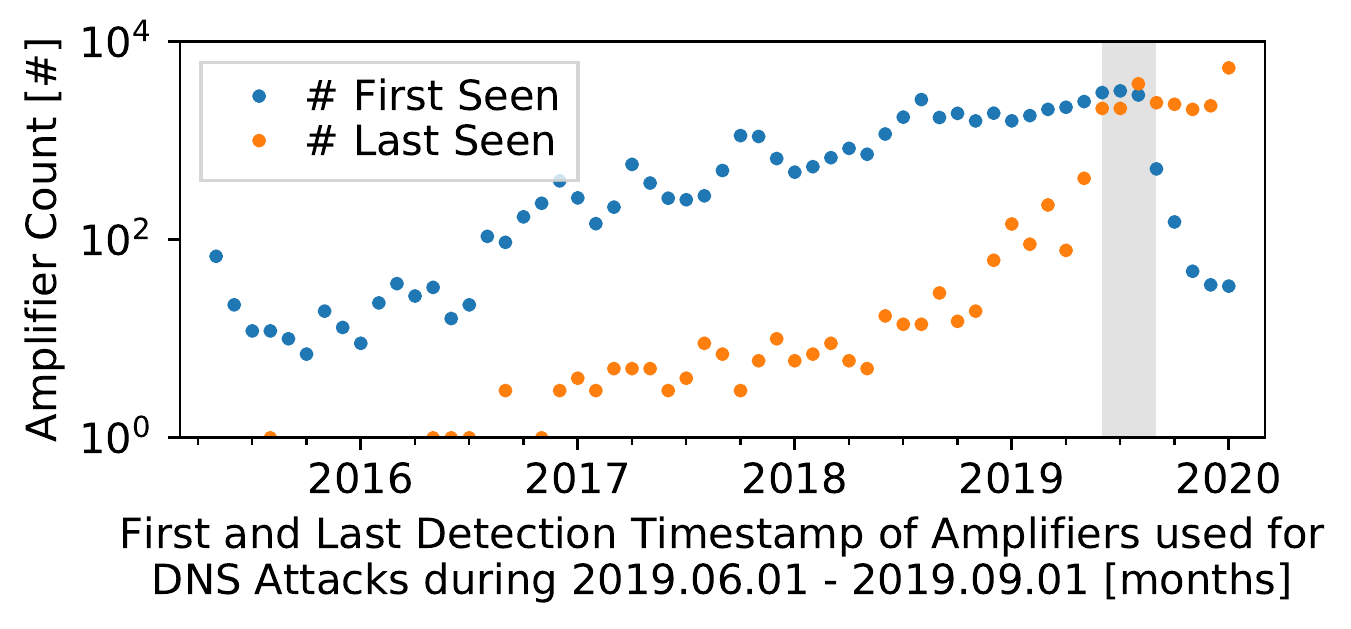}
  \caption{Number of Shodans first and last interaction with reflectors that were observed by us during the attacks at the IXP. Timerange of attacks highlighted in gray.}
  \label{fig:dns_shodan_amplifier_detection}
\end{figure}

To examine the age of an amplifier that was abused during our measurement period,
we determine its first and last successful detection by Shodan 
in \autoref{fig:dns_shodan_amplifier_detection}.  A significant number of
amplifiers was first seen during six months preceding the attack, \ie
attackers mostly use amplifiers that are not older than six months.  Also, many
amplifiers are observed for the last time during or right after our main
measurement period indicating that
\one operators change the inadvertently open state of their resolvers; or
\two the amplifiers churned because of a dynamic IP address.
Notably around 850 reflectors (2\%) appeared in attacks before discovery 
by Shodan.  This suggests that some attackers run their own scanning engines 
with a higher scan frequency or accuracy than Shodan.

At this point, our methodology allows to passively identify DNS amplifiers
as they are abused, even before other measurement efforts succeed.
Overall, we observe substantial DNS amplifier churn at the IXP but discern no
downside for attackers.  Note that we observe actual
amplifier abuse at the IXP, not the churn in amplifier reachability (which scans can
reveal).
Although the total number of abused amplifiers remains stable between attacks,
we see  on average only $45$\% of abused 
amplifiers  in subsequent $( \textrm{day}_{i},\textrm{day}_{i+1})$
pairs.
Comparing the first and last day of our three-month measurement period, only
$20$\% of amplifiers still make an appearance. This observation suggests that attackers effectively
detect---and purposefully rotate---new DNS amplifiers.

\subsection{Potential Amplification Factors}

\begin{figure}[t]
  \begin{center}
  \includegraphics[width=0.99\columnwidth]{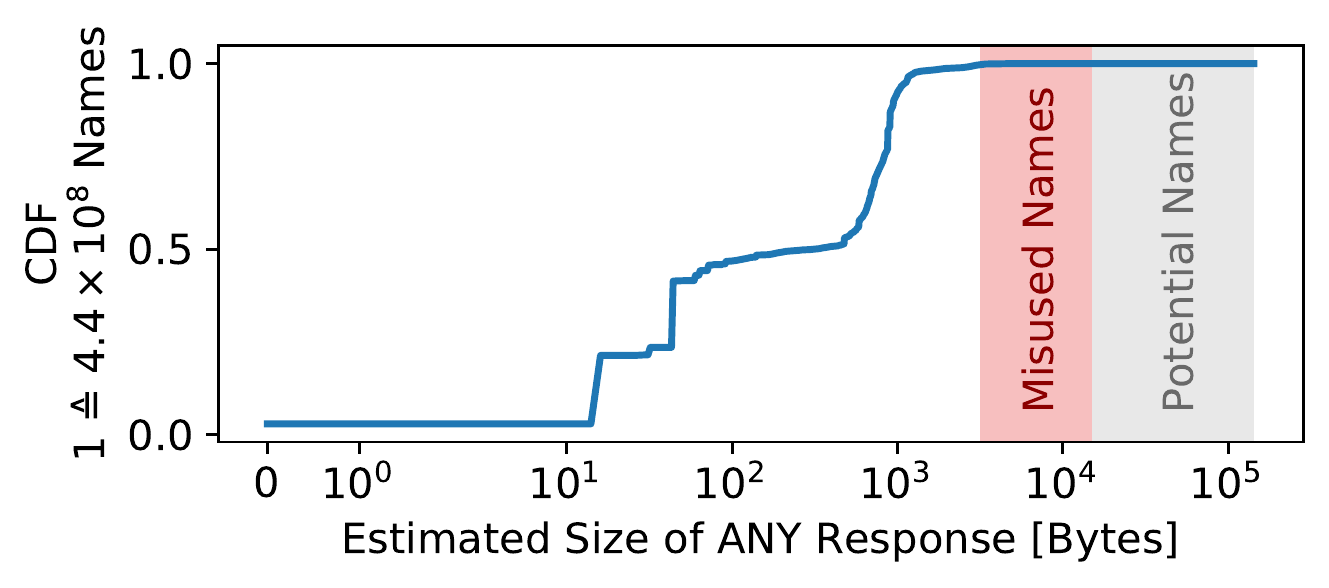}
  \caption{Estimated \texttt{ANY} response sizes for names measured by OpenINTEL. We highlight the range for currently misused names (red), and show the range of potential names (only 9048 distinct names) to increase the amplification factor~(gray).}
  \label{fig:openintel_cdf_any}
  \end{center}
\end{figure}

\paragraphNoDot{Do attackers select names that maximize amplification?}
We investigate whether attackers inquire names for maximizing amplification,
or whether there is an unused threat potential.
Using the OpenINTEL data, we estimate the response
sizes of \texttt{ANY} queries of 440~million domain names and plot the CDF, see
\autoref{fig:openintel_cdf_any}.  Please note that we calculate the 
response sizes based on the cumulative resource record sizes stored in the DNS and ignore common
software or protocol limits (4096~bytes for EDNS and 65,536~bytes for~UDP).

The names previously observed in misuse exhibit a response size highlighted in the red area.
Overall, only 9048 domains show a higher amplification factor than the highest
ranked, misused name---about 0.002\% of all names (gray area in \autoref{fig:openintel_cdf_any}).
This suggests that attackers attempt to cherry-pick names for high
amplification factors without being optimal.  Our 
 estimated largest response size is 142,855 bytes, whereas 
the largest we actually observed  is 14$\times$ smaller.

\paragraphNoDot{Can we expect larger attacks in the future?}
Frighteningly, we find that $\sim$92,000~names (0.02\%) in the OpenINTEL data set
may lead to a response size larger than 4096~bytes. Even though
EDNS~\cite{RFC-6891} recommends to not send larger replies, our measurements
reveal that the DNS infrastructure frequently  does so in practice
(see \autoref{sec:major-attack-name-fingerprinting}).

Visible DNS attack events contribute a substantial amount of DNS attack traffic to the Internet core. 
Notably, the overall attack traffic at the IXP accounts for 5\% of
the total DNS packets and 40\% of the total DNS traffic volume.  This trend becomes even
more apparent when only \texttt{ANY} traffic is considered: 68\% of \texttt{ANY}
packets and 78\% \texttt{ANY}
bytes are part of attacks.  The situation will grow worse when
attackers begin to use the names with a higher amplification factor.

\section{Discussion}
\label{sec:discussion}

\paragraphNoDot{Is the observation of the major attack entity a bias of our vantage point?}
Despite being a central element of today's Internet, most IXPs still operate locally to interconnect networks.
To verify that our observations are not a local phenomena based on our large, regional IXP, we assume that popular names are likely to be cached in the DNS.
To quantify world-wide usage of names, we apply a modified cache snooping analysis.
In a nutshell, we resolved misused names as well as a set of arbitrary names via all public resolvers and compare whether the names were cached or not (details see \autoref{apx:cache-snooping}).
We correlate the cache hits and misses with popularity of the names in the Alexa~ranking.
For reference purposes we created a name only for this study (\ie a name that was not cached before).

\begin{figure}%
  \begin{center}
  \includegraphics[width=0.99\columnwidth]{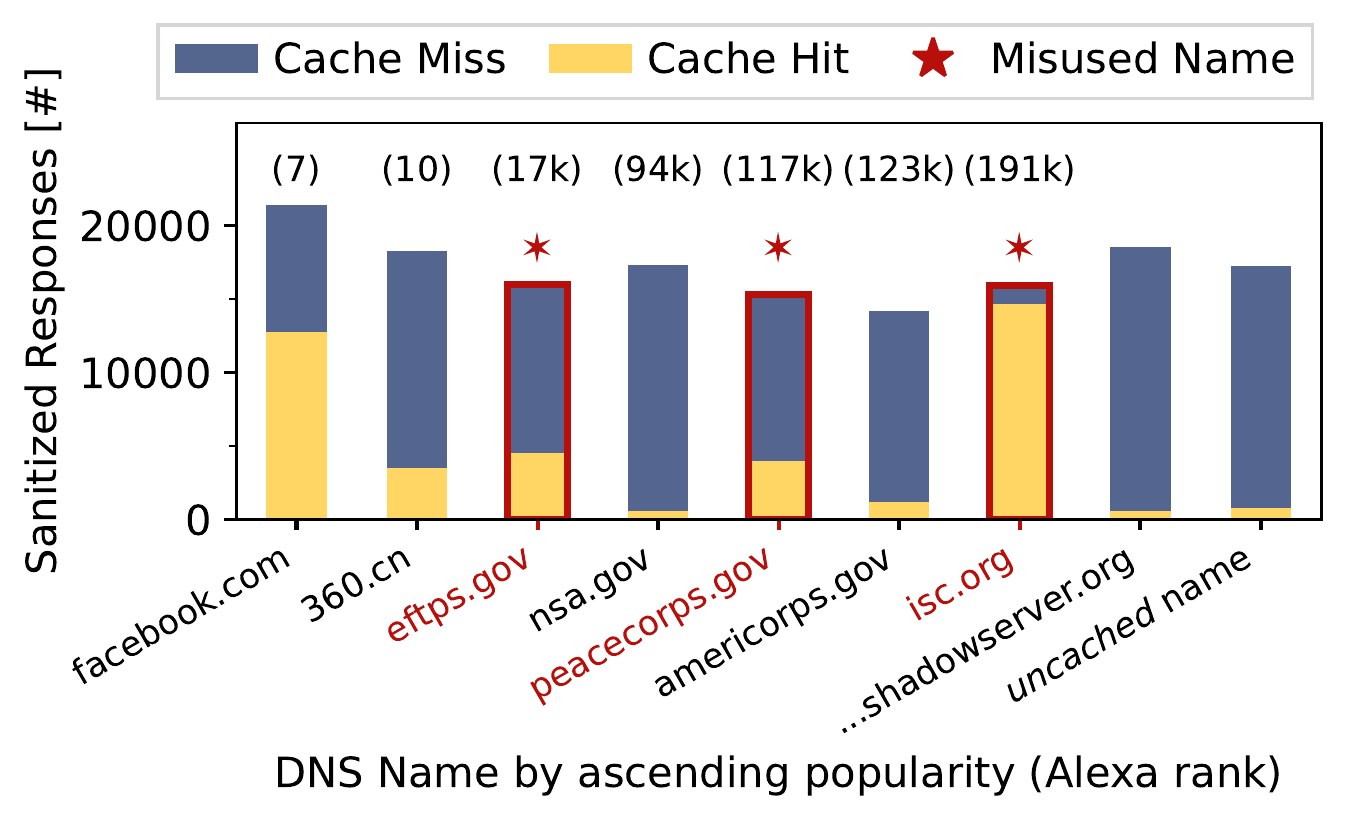}
  \caption{DNS cache hits for a set of arbitrary names and misused names. Names with a low popularity in the Alexa ranking but high cache hit rates indicate world-wide usage of the names for other reasons.}
  \label{fig:cache_snooping}
  \end{center}
\end{figure}

\autoref{fig:cache_snooping} shows that misused names (highlighted in red and with a $\star$) have a similar cache hit ratio as very popular western and eastern names, even though our misused names exhibit much lower popularity based on the Alexa ranking.
The results indicate that the misused names are resolved frequently but not because of common (Web) services.
Hence, we argue that the IXP and our methodology give insights into behavior of global scale.

Our results are further substantiated by a recent study published by an anti-DDoS provider \cite{labovitz2021nanog}.
Labovitz~\cite{labovitz2021nanog} confirms that one of the misused names identified by us (\texttt{peacecorps.gov}) has been utilized by the booter \emph{SynStresser} to perform attacks.
Also, some of our attack events correlate with publicly documented attacks \cite{rusembassy2020twitter}.

\paragraphNoDot{Would authoritative name servers provide a complete picture?}
No. 98\% of open DNS amplifiers are forwarders and not recursive resolvers~\cite{nksw-tfuco-21}.
This means that the majority of amplifiers do not communicate with authoritative servers.
Also, recursive resolvers will contact an authoritative server only when the name is not locally cached.
Cached responses, however, are common because they make DNS scalable.
TTLs may range between 1~hour up to days \cite{cache2019moura}.
We observed the impact of caching in the bi- and tri-modal distributions of attack traffic at the IXP (see \autoref{sec:trace_attackers}).
Furthermore, our data corpus includes individual resolvers that serve up to 20k~DNS amplifiers, which illustrates that caching is more likely and, thus, requests less visible at the authoritative~servers.
Hence, neither the misuse of a name nor the complete attacking infrastructure might be visible to an authoritative server.

\paragraphNoDot{What can operators do to improve the situation?}
Operators can help by configuring their authoritative nameservers or recursive resolvers to
\one~block \texttt{ANY} requests completely,
\two~respond to \texttt{ANY} requests only via TCP or with a minimal subset \cite{RFC-8482},
\three~deploy rate limiting.
Similar recommendations have been proposed for years~\cite{rvr2014dnssec}, unfortunately DNS amplifiers still exist.
Our observations suggest that those countermeasures are still helpful because an attack is based on a relatively stable set of queries.
In case advanced query patterns~\cite{nxnsattack2021afek} are issued in the future, which our method would detect, the deployment of filters that focus on names or observations across multiple resolvers are options.
As we found that some few resolvers serve a significant amount of amplifiers (\ie forwarders), educating those first will have larger impact.

To the best of our knowledge, this is the first paper that shows the adverse impact of DNSSEC key rollovers in the context of amplification attacks (see \autoref{sec:major-attack-name-fingerprinting}).
Double-signature rollovers temporarily create a second, superfluous set of signatures, which makes these names more attractive to attackers.
Operators should pay attention to misuse during rollovers for their~zones.
Overall, we recommend pre-publish rollovers which currently are best practice~\cite{pft2021dnssec}.

\paragraphNoDot{Advantages of IXPs?}
IXPs are considered to be central vantage points~\cite{benefits2013chatzis}. %
We introduced methods to leverage IXPs to shed new light on DNS amplification attacks.
We found that honeypot platforms see less compared to what was assumed before, extending recently observed trends~\cite {ddos2021kopp}.
To achieve a similar coverage compared to large, regional IXPs, honeypots require broader distribution.
What concerns us most is that honeypots are easy to detect \cite{honeypot2010wang,detection2007mukkamala,ph-vhfbt-08}, either because they deploy (for good reasons) rate limiting \cite{thomas20171000,kramer2015amppot} or they expose other features such as delays that enable fingerprinting.
Prior work clearly indicated that malware adapts and hides \cite{holz2005detecting}.
In contrast to honeypots, IXPs are native part of the Internet infrastructure.
They do not need to deploy detection schemes that expose to an attacker.
They allow for monitoring of Internet~traffic where networks intertwine, which also simplifies operational maintenance of a monitoring system.

\section{Conclusion and Outlook}
\label{sec:conclusion}
   
We studied the DNS amplification ecosystem from the Internet core, in
combination with complementary data sources.%

Our attack detection method for public peering points has enabled us to unveil
distributed inter-domain attacks.  Our results show that the DNS attack vector
is more popular than previously captured by (even distributed) honeypots, a
common vantage point in the context of reflection and amplification attacks.
We were successful in tracking a prominent attack entity and identifying
concrete attack patterns.
Our study reveals that attackers are able to detect new abusable amplifiers
quickly and reasonably change which infrastructure they abuse. At the same time,
we find that attackers could achieve higher amplification by choosing (query)
names more prudently.
especially in the case of attacks utilizing spoofing and highly variable
amplifier~sets.

Our study also reveals that operators of various US federal government domain
names break from recommended DNSSEC key rollover practices, which does not only
exacerbate the amplification potential of various \texttt{.gov} names, but
which our results can also tie to amplification attacks and attacker
decision-making.
For future work we plan to extend our methods to cover a larger number of
protocols and explore the fine-tuning of our thresholds to identify more subtle
attacks. %

\begin{acks}
We would like to thank Timm B\"ottger and Roland van Rijswijk-Deij for their insightful feedback on a previous version of this article, and Eric Osterweil and Pouyan Fotouhi Tehrani for discussions on the anatomy of DNSSEC key transitions.
We also thank our shepherd Taejoong ``Tijay'' Chung and the anonymous reviewers for their valuable comments.
This work was supported partly by the \grantsponsor{BMBF}{German Federal Ministry of Education and Research (BMBF)}{https://www.bmbf.de/} within the project \grantnum{BMBF}{PRIMEnet}.
\end{acks}

\label{lastpage}

\bibliographystyle{ACM-Reference-Format}
\bibliography{bibliography,rfcs}

\balance
\begin{appendix}
  \section{Ethics}

Our research may raise the following ethical concerns.

\paragraph{Privacy Invasion through Deep Packet Inspection}
Our IXP vantage point provides a view into application-layer payloads.  These data
are particularly sensitive as they can contain personal information, or reveal
the interests of users (\eg visited websites).
However, we do not use the data to identify or study users. %
We also present only aggregated views, eliminating the possibility for
third-parties to infer privacy-sensitive information.
Finally, we focus on attack traffic, which consists of misused query names that
do not disclose the interests of particular users.

\paragraph{Educating Attackers}
This paper presents misused query names in clear view, effectively showing the
attackers suitable names for amplification.  We argue that these names are
already extensively misused in attacks, hence publishing them will not reveal
new information.
At the same time, we identified over 9000 names that can offer higher
amplification than what we witness in practice. We will not divulge these
names.

\paragraph{Alerting the Major Attack Entity}
Releasing a DNS signature as detailed as presented in
\autoref{sec:trace_attackers} could warn the attack entity responsible for more
than half of the attacks.  We argue, however, that publishing this information
can do more good than harm as it will assist mitigation efforts by the research community.

  \section{Validation of the CCC~Honeypot Platform}
\label{apx:honeypot-convergence}

To verify that the CCC~honeypot used in this paper makes similar observations compared to previous honeypot studies, we compare various attack thresholds and analyze the convergence of our honeypot platform.
CCC infers attacks using a threshold of 5~requests per sensor with no gap of more than 900~seconds before stop replying to requests.
This is in contrast to other honeypots that set higher thresholds (\ie 100 packets and no gap of more than 3600 seconds \cite{kramer2015amppot} or 600 seconds \cite{noroozian2016gets}).
Therefore, CCC applies a more sensitive attack detection, which becomes apparent by a slightly higher number of reported DNS attacks for similar time ranges \cite{thomas20171000, kramer2015amppot}.

\begin{figure}[h]
  \begin{center}
  \includegraphics[width=0.99\columnwidth]{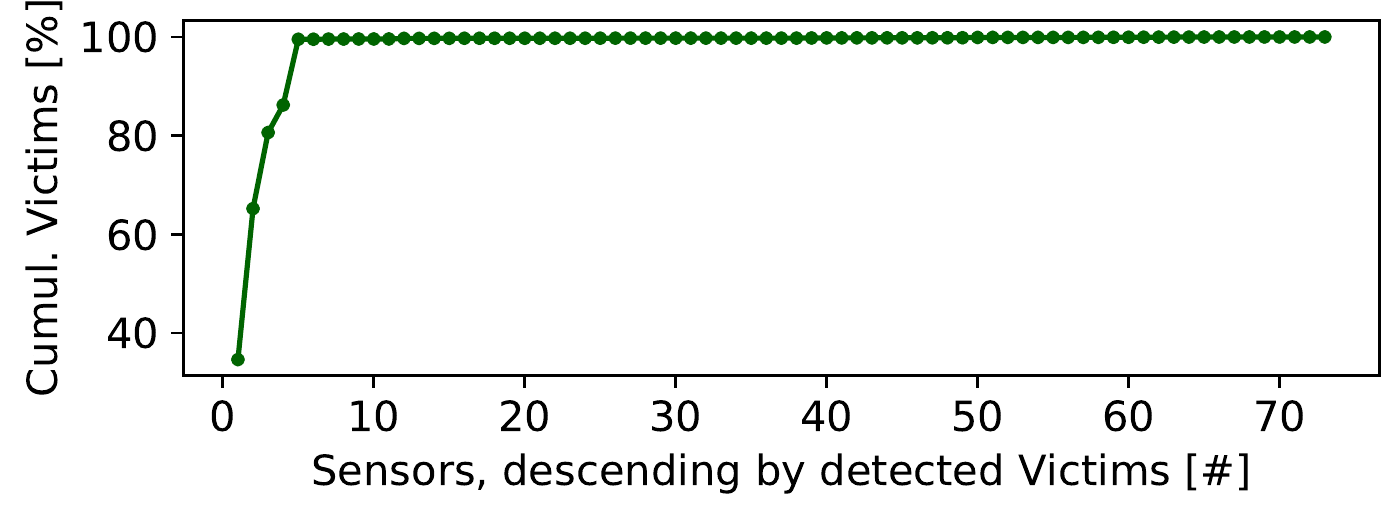}
  \caption{Honeypot convergence for the CCC platform. We observe a similar behavior compared to related projects.}
  \label{fig:honeypot_convergence}
  \end{center}
\end{figure}

A major property of honeypot platforms is the convergence of visible attacks by deploying a small number of sensors \cite{kramer2015amppot}.
We reproduce the convergence analysis of prior work~\cite{kramer2015amppot} for our data gathered at the CCC platform and make very similar observations.
99.5\% of victims are already visible with only 5 sensors, see \autoref{fig:honeypot_convergence}.
However, we require 50 sensors to cover 99.9\% of victims due to a long-tail distribution, 
It is worth noting that the CCC platform is assumed to capture most DNS attacks (between  85.1\%  and  96.6\%) on the basis of a capture-recapture statistical   technique \cite{thomas20171000}.

Overall, these results suggest that our honeypot platform behaves similar to related projects.
The sensitive thresholds and convergence behavior suggest the observation of all DNS attacks, which we refute in \autoref{sec:disjoint_attacks}.

\section{Cache Snooping to Check Name Popularity}
\label{apx:cache-snooping}
To verify whether a name is frequently resolved globally, we use a modified cache snooping (CS) analysis.
CS exploits the fact that popular names remain in DNS caches.
Cached responses are identified by TTLs which are smaller than the default TTL defined by the authoritative nameservers.
CS has been used to scrutinize the caching behavior at large public DNS resolvers \cite{trufflehunter2020randall}.
Today, there is relatively little guidance backed by research about how to set TTLs, so operators usually reuse the same TTLs (\eg 5~minutes, 1~hour, 1~day) \cite{cache2019moura}, which makes this analysis easier.

\paragraph{Phase 1: Identifying DNS~Resolvers}
We perform a scan of the complete public IPv4 address space and search for DNS~amplifiers.
Simply initiating DNS queries to all potential amplifiers and checking whether the DNS~TTLs comply with default TTLs does not yield accurate results.
This is due to the common DNS deployment, in which a DNS~forwarder uses a recursive resolver and thus inherits current TTLs from this resolver.
Hence, we first need to exclude forwarders from CS. %

To identify DNS~resolvers, we operate our own name and authoritative DNS~server that responds with an \texttt{A}~record set to the IP address of the resolver that directly queries our authoritative nameserver. 
By comparing both the IP~address of the \texttt{A} record and target with the source IP~address of the respond, we can distinguish resolvers (addresses match) and forwarders (addresses differ), details see \cite{nksw-tfuco-21}.
This method allows for fast scanning with pre-built DNS queries and also limits the traffic at our authoritative nameserver since forwarders using the same resolver will return a cached entry.
The relation between forwarders and resolvers has been measured before, but the previous methodologies~\cite{anagnostopoulos2013dns, client2013schomp, inferring2020korczynski} embed the IP address of each target into the subdomain.
This embedding requires the analysis of queries at the authoritative nameserver, which impedes reproducibility.

\paragraph{Phase 2: Assessing Name Popularity}
After isolating the resolvers, we now can initiate a CS \texttt{ANY} scan to find uncommon cache activities.
The reasoning here is that misused names are uncommonly often present in caches although being not popular, as measured by \eg the Alexa Rank.
We find recently misused names with the help of our long-term monitoring tool.
We sanitize responses by removing
\one answers with erroneous flags and codes (\eg rcode \texttt{REFUSED}),
\two responses from obvious DNS manipulators, (\ie sources that change the TTLs or \texttt{A} records),
\three duplicate responses from a single source.
Then, we classify a response as a cache miss if all answer resource records contain a default TTL, a cache hit otherwise.

We focused our analysis on similarly popular \texttt{.gov} names.
Please note that \texttt{americorps.gov} also has a larger (+25\%) maximum TTL than \texttt{peacecorps.gov}, so it would be expected to produce more cache hits.

We utilize two anchor names to verify the correctness of this measurement.
First, we reuse a name from Shadowserver which has well-documented, daily scanning times and TTLs.
We initiate our scans \emph{after} the daily expiration time to showcase correct cache evictions.
Second, right before our CS scan, we create a completely new name, which should produce cache misses only.
Still, the anchor names reveal a small amount of cache hits.
We consider these cache hits to be the error rate of our measurements.
We assume mutual resolver caches and DNS optimizers responsible for these errors.

\end{appendix}

\end{document}